\documentclass[twocolumn, usenatbib]{mnras}
\pdfoutput=1 
\usepackage{amsmath,amstext}
\usepackage{appendix}
\usepackage[T1]{fontenc}
\usepackage{ae,aecompl}
\usepackage[figure,figure*]{hypcap}
\usepackage[graphicx]{realboxes}
\usepackage{amssymb}	
\usepackage[caption=false]{subfig}
\usepackage{tikz}
\newcommand{\angstrom}{\textup{\AA}}
\newcommand{\fluxunits}{$10^{-17}$ erg/s/cm$^2$/\AA}

\usepackage{soul,xcolor}
\setstcolor{red}

\makeatletter
\renewcommand*{\@fnsymbol}[1]{\ifcase#1\or ^\dagger\or ^{\star} \else\@ctrerr\fi}
\makeatother
\title[eBOSS: Strong Galaxy Lens Candidates]{The Completed SDSS-IV extended Baryon Oscillation Spectroscopic Survey: A Catalogue of Strong Galaxy-Galaxy Lens Candidates}

\author[Talbot and Brownstein et al.]{
\parbox{\textwidth}{Michael S.\,Talbot,$^{1}$\thanks{E-mail: \texttt{\href{mailto:michaeltalbot@astro.utah.edu}{michaeltalbot@astro.utah.edu}}}
Joel R.\,Brownstein,$^{1}$\thanks{E-mail: \texttt{\href{mailto:joelbrownstein@physics.utah.edu}{joelbrownstein@physics.utah.edu}}}
Kyle S.\,Dawson,$^{1}$
Jean-Paul Kneib,$^{2}$
Julian Bautista$^{3}$
\\}
\\
$^{1}$Department of Physics and Astronomy, University of Utah, 115 S. 1400 E., Salt Lake City, UT 84112, USA \\
$^{2}$Institute of Physics, Laboratory of Astrophysics, Ecole Polytechnique Fédérale de Lausanne (EPFL), Observatoire de Sauverny, \\ \phantom{$^{2}$}1290 Versoix, Switzerland \\
$^{3}$Institute of Cosmology \& Gravitation, University of Portsmouth, Dennis Sciama Building, Portsmouth, PO1 3FX, UK
}

\date{Accepted for publication by MNRAS 27-Jan-2021. Received 27-Jul-2020; in original form 17-Jul-2020.}
\pubyear{2021}
\begin{document}
\maketitle
\begin{abstract}

We spectroscopically detected 838 likely, 448 probable, and 265 possible strong lens candidates within $\approx2$ million galaxy spectra contained within the extended Baryon Oscillation Spectroscopic Survey (eBOSS) from the sixteenth data release (DR16) of the Sloan Digital Sky Survey (SDSS). We apply the spectroscopic detection method of the Baryon Oscillation Spectroscopic Survey (BOSS) Emission-Line Lens Survey (BELLS) and add Gaussian fit information, grading, additional inspection observables, and additional inspection methods to improve our selection method. We observed 477 candidates with lensing evidence within low-resolution images from both the Legacy survey of SDSS-I/II and the DESI Legacy survey, which is $12\%$ higher than the percentage of BELLS candidates observed with similar lensing evidence. Our search within the latest and improved reductions of the BOSS survey yielded a $20\%$ increase in the number of lens candidates expected from searching all BOSS and eBOSS galaxies.  The distribution of target and background redshifts of our candidates is similar to the candidates and confirmed lenses within the BELLS observations. We present our Spectroscopic Identification of Lensing Objects (SILO) candidates in a value-added catalogue (VAC) in SDSS DR16. The examination of these lens candidates in follow-up high-resolution imaging may yield more than twice the lenses found in previous spectroscopic detection surveys within SDSS, which would extend the results of previous lens surveys within SDSS to higher redshifts, constrain models of mass structures in spiral galaxies, and test if including the identification of possible lensing features within low-resolution images has merit to spectroscopic detection programs.

\end{abstract}

\section{Introduction}
\label{section:introduction}

\subsection{The Spectroscopic Identification of Lensing Objects}\label{subsection:introduction.objective}
 
We are interested in the many ways that galaxy-scale strong gravitational lensing enables the measurement of the galactic mass distribution, and to decompose the dark matter distribution to explore the effect on galaxy formation and evolution, and to use this data to refine the standard model of cosmology to further our understanding of the distribution of matter in the Universe. We are also interested in how precise gravitational lensing models may test general relativity and gravity theories.

Obtaining large samples of confirmed galaxy-scale lenses with Einstein radii distributed across the radial extent of the galactic profiles would enable us to probe further into the baryon-dominated galaxy core and also out farther into the dark matter dominated halo. An effective way to acquire this needed spread, commonly parameterized by the ratio of Einstein radii to the effective radius, is to search large surveys for galaxy-galaxy strong lenses across several ranges of redshift, which also enables us to measure distance and time-dependent trends in the galaxy structure, to study galaxy formation and evolution, and cosmology-scale properties.

Our overall project objective is to discover and confirm $o(10^3)$ grade-A strong gravitational lenses within the Sloan Digital Sky Survey (SDSS) including the SDSS-I/II Legacy~\citep{2001ASPC..238..269L} survey, the Baryon Oscillation Spectroscopic Survey~\citep[BOSS;][]{2013AJ....145...10D} from SDSS-III, the extended Baryon Oscillation Spectroscopic Survey~\citep[eBOSS;][]{Dawson_2016}, and the Mapping Nearby Galaxies at APO~\citep[MaNGA;][]{2015ApJ...798....7B} survey of the current phase of the Sloan Digital Sky Survey~\citep[SDSS-IV;][]{2017AJ....154...28B}. Such a catalogue would be the first of its kind, and the current version is available to the public, included in the Sixteenth Data Release (DR16) of the SDSS-IV~\citep{2020ApJS..249....3A}. We provide a brief overview of galaxy-galaxy strong gravitational lensing in Section~\ref{subsection:introduction.overview}, and review the progress of the spectroscopic detection method in SDSS in Section~\ref{subsection:introduction.detection}.

\subsection{Galaxy-Galaxy Strong Gravitational Lensing}\label{subsection:introduction.overview}

The gravitational bending of the light from a distant source galaxy as it passes near the influential mass of the lens galaxy can generate multiple images of the source at the position of an observer along the line-of-sight (LOS) of the converging rays from the images. These strong galaxy-scale gravitational lenses will magnify at least one of the warped source images. In addition, grade-A lenses clearly demonstrate multiple source images that can be modeled to provide precise values of the Einstein radii, which are used as mass probes into individual galaxies and cosmological parameters, including the expansion rate of the Universe that influences the difference in the arrival time of light between multiple source images. These properties of galaxy-galaxy strong gravitational lenses have been used to constrain cosmology and galaxy evolution in a variety of ways, such as:

\begin{itemize}
\item Using the lens as a telescope to examine magnified spacial or spectral features of galaxies (which may be unresolved without lensing) to study source properties such as star formation, stellar populations, size and compactness of the source, and classify source into categories such as a compact blue dwarf, a low-metallicity star-bursting dwarf galaxy, or an early-type galaxy~\citep{Marshall_2007, Cabanac_2008, Brammer_2012, Oldham_2016, 2017arXiv170808854C}. 
\item Fitting a galaxy mass model to how the lens is bending the source light to measure the total mass scale and mass distribution of the lens, determined by the best-fit model. These measurements can then be combined with the lens' light profile to determine the total mass to light ratio. Fitting stellar (and even gas) mass models to the light from the lens enable the lens mass to be separated into the contributions from baryons and dark matter. All of these measured components can be used to test and constrain models of the mass structure of galaxies, which impact galaxy formation and evolution models~\citep{1995ApJ...445..559K, Gavazzi_2008, 2012MNRAS.424.2800L, Nightingale_2019}.
\item Combining lensing with rotation curves or other dynamic and kinematic data to significantly constrain measurement uncertainties and enable the dissection of the mass distribution into the different stellar, gaseous, morphological, and dark components of galaxies~\citep{2011MNRAS.417.1621D, 2011MNRAS.417.3000S, 2012MNRAS.423.1073B}.
\item Measuring the mass distributions of a large set of lenses (which can be combined with dynamical mass measurements) across cosmological redshifts enables a probe into trends within galactic mass profiles across distance and time. This information is useful to constrain time and distance independent or  dependent mass components of galaxy structure, formation, and evolution models~\citep{1998ApJ...495..157K, 1998ApJ...509..561K, Koopmans_2006, 2008ApJ...684..248B, 2009MNRAS.399...21B, Koo++09, 2010ApJ...724..511A, Sonnenfeld_2013, 2015ApJ...803...71S, Cao_2016, Li_2018, mukherjee2019seagleii, Sonnenfeld_2019}.
\item Probing possible indications of dark matter halos perturbing the source light along the line-of-sight (LOS)~\citep{2017arXiv171005029D}.
\item Probing for dark substructures near or within the lens using methods such as gravitational imaging~\citep{10.1111/j.1365-2966.2005.09523.x, 2010MNRAS.408.1969V, 2014MNRAS.442.2017V, 2016ApJ...823...37H}.
\item Detecting dark satellite galaxies perturbing the path of the source light~\citep{2012Natur.481..341V}.
\item Using the time-difference measured between flux-varying source images to better constrain angular diameter distances and components in the lens model~\citep{2015JCAP...11..033J}.
\item Testing the cosmological constant \(\Lambda\) in the standard model of cosmology by comparing lens statistics to derived expectations~\citep{1990MNRAS.246P..24F}.
\item Providing insights that help refine concepts related to Cold Dark Matter (\(\Lambda\)CDM) models~\citep{ benson2019normalization}.
\item Constraining the dark energy equation of state and other components of cosmography by examining rare lenses with two sources~\citep{Gavazzi_2008, 2012MNRAS.424.2864C, Linder_2016}.
\end{itemize}

Most of the research efforts listed above utilize large samples of lenses to either observe trends in galactic mass profiles or to search for unusual lens systems within the sample. Obtaining large samples of galaxy-scale strong gravitational lenses is difficult since a source galaxy has to be sufficiently distant from the lens galaxy with an alignment within arc-seconds of the line-of-sight from the lens galaxy to the distant observer. Most of these chance alignments are unlikely to be detected within images since the dominating foreground light of the lens galaxy often conceals most or all of the background source light. It is also difficult to differentiate unclear features of a source from either an un-lensed background galaxy, multiple background galaxies, satellite galaxies, structure in the lens, improper image reduction, or other causes of false-positive features.

\subsection{Spectroscopic Detection within SDSS}\label{subsection:introduction.detection}

One of the most successfully tested solutions to obtain a large sample of confirmed galaxy-scale lenses is to spectroscopically detect background emission-lines from star-forming galaxies behind the target galaxy within a sample on the order of a million galaxy spectra, followed by precisely subtracting the foreground lens light in follow-up high-resolution images to confirm the source.

The Sloan Digital Sky Survey~\citep[SDSS;][]{2000AJ....120.1579Y} contains the largest fraction of confirmed lenses, with at least 145 grade-A strong galaxy-scale lenses confirmed from at least 311 lens candidates spectroscopically detected by the Sloan Lens ACS~\citep[SLACS;][]{2006ApJ...638..703B} survey, the Sloan WFC Edge-on Late-type Lens Survey~\citep[SWELLS;][]{2011MNRAS.417.1601T}, and the SLACS for the Masses~\citep[S4TM;][]{2015ApJ...803...71S} survey within the $\approx1$ million galaxy spectra (z < 0.55) contained in the Legacy survey of SDSS-I/II. These lens surveys verified the source images after subtracting the lens light in follow-up images obtained either by the Hubble space telescope (HST) or Keck-II.

Moving out to greater target redshifts in the SDSS-III/IV era, the BOSS and eBOSS surveys used the BOSS spectrograph~\citep{2006AJ....131.2332G, 2013AJ....146...32S} to provide more than twice as many galaxy spectra up to twice the redshift (z <= 1.1) of the galaxy sample contained within the Legacy survey of SDSS-I/II. The BOSS Emission-Line Lens Survey~\citep[BELLS;][]{2012ApJ...744...41B}, based on the spectroscopic detection method of the SLACS survey, yielded 25 grade-A strong galaxy-galaxy scale gravitational lenses from 45 lens candidates at the start of the BOSS survey following the Ninth Data Release (DR9), which contained only $\frac{1}{17}$ of the galaxy spectra of the now completed BOSS and eBOSS surveys in DR16. The BELLS lenses were confirmed by HST Program 12209, including precise lens models constructed after subtracting the foreground lens light in the follow-up images.

Several other lensing surveys have successfully identified a fraction of the spectroscopically detectable candidates within the BOSS survey. The BELLS GALaxy-Ly$\alpha$ EmitteR sYstems~\citep[BELLS GALLERY;][]{2016ApJ...824...86S} survey found 21 strong gravitational lens candidates (17 confirmed) with candidate Lyman-alpha emitters (LAEs) sources by searching for Lyman-alpha emissions behind the target. The~\citep[SuGOHI;][]{Sonnenfeld_2017} survey has developed a combined image and spectrum search method that included a spectroscopic search of $~\frac{1}{20}$ of the BOSS targets. Thus the remaining $~\frac{15}{17}$ of the BOSS and eBOSS targets were not scanned for the emission-lines searched within spectroscopic detection surveys prior to the SuGOHI survey (see table \ref{table:emline}).

This work extends the Spectroscopic Identification of Lensing Objects~\citep[SILO;][]{10.1093/mnras/sty653} program to complete the search for galaxy-galaxy lens candidates within the latest reductions of the galaxy spectra contained within SDSS-IV Data Release 16 of the completed BOSS and eBOSS surveys. SILO is the current evolution of the SDSS spectroscopic detection code that displays significantly more information on the quality of the candidate emission-lines and possible sources of contamination (see Sections~\ref{subsection:method.gauss} and~\ref{subsection:method.inspection}) by means of a manual inspection (Python) interface.

We also inspected low-resolution images of the target to assess if interference from foreground objects may impact the signal and if lensing evidence is observed within the images (see Section~\ref{subsection:method.imaging}). We created a grading scheme which is calibrated to assign a grade of A- or above to candidates with emission-lines as clearly defined as the confirmed emission-lines observed within the BELLS survey. We also assign each candidate a total grade based upon the assurances observed within both the signal quality and evidence of lensing.

This paper is organised as follows. The expected number of lenses and lens candidates that SILO can detect within the BOSS and eBOSS surveys are described in Section~\ref{section:expectation}. The spectroscopic detection method is presented in Section~\ref{section:method}, consisting of foreground spectrum modeling and subtraction (Section~\ref{subsection:method.galaxy_template}), background emission-line detection (Section~\ref{subsection:method.emline}), Gaussian modeling of the candidate emission-lines of the candidate background galaxy (Section~\ref{subsection:method.gauss}), candidate cuts with respect to likely lensed systems (Section~\ref{subsection:method.prerefine}), inspection method of the spectra (Section~\ref{subsection:method.inspection}), and inspection of low-resolution images (Section~\ref{subsection:method.imaging}). We present a summary of our findings in Section~\ref{section:data}, consisting of the counts of detected candidates and examples of lens candidates (Section~\ref{subsection:results.spectracount}), counts of the types of features observed in low-resolution images that may be related to lensing (Section~\ref{subsection:results.imagecount}), and comparison of SILO lens candidates to the BELLS survey (Section~\ref{subsection:results.bells}). We describe the publicly available value-added catalogue (VAC) containing the BOSS and eBOSS lens candidates in Section~\ref{section:vac}. We provide a summary and discuss some examples of how follow-up imaging of these lens candidates can extend the results of previous lens surveys, enable new lensing research, and provide opportunities to bridge information from spectroscopic detection with low-resolution images in future discovery programs (Section~\ref{section:conclusion}).

\section{Expected Number of Highly Assuring Lens Candidates and Confirmable Grade-A Lenses}\label{section:expectation}

Since the SILO and BELLS discovery programs follow a common detection method (see Section~\ref{section:method}) to find lenses within SDSS-III BOSS data, we can approximate the number of BOSS lens candidates SILO should detect by scaling up the number of lens candidates found in the BELLS survey by the fractional increase of SDSS-III BOSS galaxies searched by SILO over the BELLS survey, which was limited to DR10. The $\approx1.5$ million galaxy spectra of the improved (see reduction improvements in~\citet{Ahn_2014, Alam_2015, 2016arXiv160802013S, 2017arXiv170709322A}) and final DR16 data reduction pipeline ($v5\_13\_0$) contains thirteen times the luminous-red-galaxy (LRG) spectra available to the BELLS. If we neglect the additional inspection methods, additional spectral observables, additional image observables, and improved pipeline of the BOSS data, then we expect SILO to result in spectroscopically selecting $\approx13 * 45 = 585$ lens candidates within the BOSS data, with a similar quality of signal to the BELLS candidates. Since 25 ($56\%$) of the lens candidates from the BELLS survey w  ere confirmed as grade-A lenses, we can extrapolate that approximately $\approx13 * 25 = 325$ of SILO lens candidates would be confirmed as grade-A lenses, under the same assumptions as above. We do not remove any previously discovered candidates from this approximation since previously discovered candidates are included in the DR16 results, in order to provide a self-consistent and complete value-added catalogue to the public, as a comparative resource.

In addition to the SDSS-III BOSS data, DR16 contains 0.5 million galaxy spectra targetted by the SDSS-IV extended BOSS (eBOSS) survey, which is $\frac{1}{4}$ the number of galaxy spectra contained within the BOSS survey. Thus the total galaxy spectra contained in DR16, which includes all of the BOSS and eBOSS targets, and searched by SILO, is a factor of seventeen more than the galaxy spectra searched by the BELLS.
However, several differences between the BOSS and eBOSS galaxies imply that the SILO detection rate for eBOSS targets will be significantly reduced by the following limiting factors:
\begin{enumerate}
\item The target selection of eBOSS galaxies is extended to higher redshifts by \(+0.4\). The higher redshifts of the eBOSS targets require the redshift of any background source to typically be higher than in BOSS targets for strong lensing to occur. Thus the search window between the lower redshift limit of candidate eBOSS sources and the maximum limit of the spectrograph is typically smaller for eBOSS targets than for BOSS targets.
\item Sixty per cent of the eBOSS targets are emission-line-galaxies (ELG), which are typically less massive than the luminous red galaxies (LRG) from the BOSS survey. Thus the fraction of background sources whose signal was magnified by lensing above our detection threshold will be less for eBOSS galaxies in comparison to BOSS galaxies since the lower ELG masses will decrease the magnification power of the lens, and reduce the Einstein radius of lenses within the eBOSS survey.
\end{enumerate}

It is important to note that a typically smaller Einstein radius for eBOSS candidates may enable more sources to be found within the \(1\arcsec\) solid angle of the spectroscopic fibre. However, blurring from seeing mitigates this advantage since sources beyond the solid angle of the fibre can be ineffectively detected~\citep{2012ApJ...753....4A}. Thus the significant differences listed above imply that SILO should find less than $585*\frac{1}{4}=146$ lens candidates from eBOSS galaxies with quality of signal similar to that of the BOSS candidates, though it is reasonable to speculate that we will select on the order of a hundred eBOSS candidates. Thus the sum of our approximations yields a total expectation of $\approx700$ SILO lens candidates with quality of signal approximate to the BELLS candidates, of which $\approx325 + \approx50=\approx375$ can be expected to be confirmed as grade-A lenses.

We emphasise that the actual yield of candidates from SILO should be greater than the above estimate, since we used the improved final ($v5\_13\_0$) data reductions of the BOSS and eBOSS spectra released with DR16, and included additional observables in the spectroscopic lens detection method. We also emphasise these estimations apply only to candidates with quality of signal approximate to the BELLS candidates, and thus the expected number of lens candidates may be better treated as a lower bound to the number of lens candidates and grade-A lenses that can be found by SILO. Results from the SLACS survey demonstrate that the spectroscopic detection rate is approximately 1 in 1000 galaxy targets within the Legacy survey of SDSS-I/II~\citep{2006ApJ...638..703B, 2009ApJ...694..924M}. Since the spectroscopic detection window between the Legacy and BOSS surveys remained approximately the same, we can speculate that the number of spectroscopically detectable lenses within BOSS and eBOSS is within an order of magnitude of 2,000 strong graviational lenses.
\section{Spectroscopic Candidate Selection}\label{section:method}

The approach of our spectroscopic detection method is to search for background emission-lines within the solid angle of the fibre~\citep[see also][]{1996MNRAS.278..139W, 2000ASPC..195...94H, 2005MNRAS.363.1369W, 2006MNRAS.369.1521W}. We first summarise our spectroscopic discovery method and then proceed to describe each step. We searched for strong galaxy-galaxy gravitational lenses within the co-added galaxy spectra contained within the completed BOSS and eBOSS surveys within the SDSS DR16 dataset. Our selection method is based upon the spectroscopic detection method, refinement cuts, and manual inspection method of the BELLS program, which were extended from the methods of the SLACS survey. The spectroscopic detection method fits a multi-component galaxy template model to the foreground flux, then removes the flux according to the model, and then searches the residual flux for higher redshift, high signal-to-noise (SN) gas emission-lines of a star-forming galaxy that is sufficiently distant behind the foreground galaxy for strong lensing to likely occur.  Since the fibre radius, exposure times, and the BOSS spectrograph are unchanged between the SDSS-III and the current search of SDSS-IV, our application of the BELLS discovery method recovered a similar signal-to-noise (SN) for the emission-lines of the candidates from the BELLS detections.

We add best-fit Gaussian models to the detected background emission-lines, additional observables of the spectra, low-resolution images of the target, and additional inspection methods to the BELLS inspection process. Detections that are likely sky, target emission-lines, misidentified background emission-lines, are unlikely to originate from background galaxies, or are unlikely to be strongly lensed are rejected in the refinement cuts, manual inspection of the spectra, and image inspection processes. The solid angle of the 1\arcsec fibre, which is ineffectively extended by seeing~\citep{2012ApJ...753....4A}, limits the LOS of the candidate background galaxy to near or often within the strong lensing regime, which enables us to promote the candidate background galaxies to strong galaxy-galaxy scale gravitational lens candidates as has been done in previous spectroscopic detection surveys within SDSS.

\subsection{Foreground Galaxy Subtraction}\label{subsection:method.galaxy_template}

We model and subtract the spectra of the foreground galaxy using the SILO code, which includes a python re-implementation of the IDL algorithms common to SLACS, SWELLS, S4TM, BELLS and BELLS GALLERY.  We first mask the contaminated parts of the co-added spectra using the \texttt{ANDMASK} supplied for each target from the BOSS and eBOSS surveys. We then construct a principal component analysis (PCA) model of the foreground galaxy spectra using the precise spectroscopic redshift obtained by the \texttt{spec1d}~\citep[][]{2012AJ....144..144B} data reduction pipeline using a basis of 7 PCA eigenspectra derived from SDSS galaxies.  Finally, we use the constructed galaxy model to subtract the light of the foreground galaxy, which produces a flattened residual flux that is dominated by noise.

\subsection{Background Emission-Line Detection}\label{subsection:method.emline}

The primary objective of our spectroscopic detection method is to find emission-lines from a galaxy far behind the target, which are caused when large, hot, and newly formed main sequence stars (i.e., such as O, B, and A stars) heat and photo-ionise the surrounding gas into emission. Since the background emission-lines are not removed in foreground subtraction, these emission-lines can appear as high signal-to-noise (SN) features within the residual flux. We compute the SN of the spectra and then search for:
\begin{enumerate}
\item Two or more of the emission-lines listed in Table~\ref{table:emline} with an SN $\geq$ 4.
\item An $[O\,\textsc{ii}]$(b, a) doublet with an SN $\geq$ 6.
\end{enumerate}
This yields a precise redshift for the candidate background source galaxy.  
 
\begin{table}
\caption{\label{table:emline}Emission-lines searched for within the BOSS and eBOSS surveys. Column 1 lists the name of each emission-line, while Column 2 lists the restframe vacuum wavelength of each emission-line. Column 3 lists the maximum redshift each emission-line can be detected within the BOSS spectrograph.}
\centering
\begin{tabular}{ c c c }
\hline
\hline
{Emission} & {Restframe} & {\(z_{\mathrm{max}}\)} \\
{Line} & {Wavelength [\AA]} & \\
{\scriptsize (1)} & {\scriptsize (2)} & {\scriptsize (3)} \\
\hline
$[O\,\textsc{ii}]$b & 3727.09 & 1.78 \\
$[O\,\textsc{ii}]$a & 3729.88 & 1.78 \\
H${\delta}$ & 4102.89 & 1.52 \\
H${\gamma}$ & 4341.68 & 1.38 \\
H${\beta}$ & 4862.68 & 1.13 \\
$[O\,\textsc{iii}]$b & 4960.30 & 1.09 \\
$[O\,\textsc{iii}]$a & 5008.24 & 1.07 \\
$[N\,\textsc{ii}]$b & 6549.86 & 0.58 \\
H${\alpha}$ & 6564.61 & 0.58 \\
$[N\,\textsc{ii}]$a & 6585.27 & 0.57 \\
$[S\,\textsc{ii}]$b & 6718.29 & 0.54 \\
$[S\,\textsc{ii}]$a & 6732.68 & 0.54 \\

\hline
\end{tabular}
\end{table}

The signal-to-noise thresholds are identical to the SLACS and BELLS surveys and are selected to maximise the detection of lensed galaxies while preventing false-positives from dominating the catalogue of candidates. The first and second search modes are known as ``multi-line'' and ``single-line'' detections, respectively. We require the redshift of these detections to be at least \(0.05\) behind the target as also required within the BELLS survey. This automatic search of DR16 produced 76,889 single-line detections and 49,732 multi-line detections.

\subsection{Fitting Gaussian Models to the Detections}\label{subsection:method.gauss}
We added Gaussian fit information to our candidate selection method to highlight more promising or questionable features in the detections and observe global trends in the signal (see Section \ref{subsection:method.inspection}). We first fit all candidate background emission-lines to a single Gaussian by minimising the $\tilde{\chi}^2$ of the fit. The inverse-variances used in the $\tilde{\chi}^2$ fit enabled us to automatically reject bad data inherent to the data reduction pipeline. We also fit the $[O\,\textsc{ii}]$(b, a) candidate emission-lines to a $[O\,\textsc{ii}]$(b, a) double Gaussian with a fixed 1:1 amplitude ratio and a redshift-dependent separation between the modelled emission lines. These constraints enabled us to use the reduced $\tilde{\chi}^2$ of the fits within three sigma of the model (including the emission-line spacing for the double Gaussian model) to infer whether the following constraints are met: 
\begin{enumerate}
\item The candidate $[O\,\textsc{ii}]$(b, a) doublet approximately matches the $[O\,\textsc{ii}]$(b, a) pattern expected from a star-forming galaxy.
\item The shape of the candidate $[O\,\textsc{ii}]$(b, a) doublet yields smaller errors of the fit to an $[O\,\textsc{ii}]$(b, a) double Gaussian than a single Gaussian.
\end{enumerate}

In contrast, these constraints do not enable the variance in the observed amplitudes of the candidate $[O\,\textsc{ii}]$(b, a) double Gaussian to be fitted individually. Thus while these constraints are ideal for comparing the quality of the fit with the expected pattern of an $[O\,\textsc{ii}]$(b, a) doublet, we emphasise that the observed details in the features of candidate $[O\,\textsc{ii}]$(b, a) doublets can be misrepresented by the fit. It is also important to note that we fit the candidate emission-lines as observed, which are not re-scaled into the candidate rest-frame of the emission. An explanation of the optimisation method to perform the fit is presented in Appendix \ref{appendix:gaussfits}.

The best fits Gaussian parameters are included in the value-added catalogue (VAC, see Section \ref{section:vac}) to provide users with an indicator of the quality of the expected patterns demonstrated by the emission-line(s) within the SILO candidates. We also provide an approximation of the uncertainty in the parameters by computing the difference between the best-fit parameters and the 16 and 84 quartiles of each parameter evaluated from 10,000 samples generated by Monte-Carlo simulations. The Monte-Carlo samples were generated by adding error-scaled Gaussian noise to the residual flux and performing the fit. Figure~\ref{figure.gauss} demonstrates the quality of the $[O\,\textsc{ii}](b, a)$ doublet Gaussian fit to the $[O\,\textsc{ii}](b, a)$ candidates.

\newcommand{\gausslinecaption}{Histogram demonstrating the quality of the $[O\,\textsc{ii}]$(b, a) doublet Gaussian fit to the candidate $[O\,\textsc{ii}]$(b, a) doublets. Candidates are binned by the reduced $\tilde{\chi}^2$ of the fit evaluated within three $\sigma_{model}$ of the emission-lines. The blue patch pattern and solid red pattern represent which portion of each bin count is found by single-line or multi-line search modes, respectively. The reduced $\tilde{\chi}^2$ of the fit is less than three for $94\%$ of the candidate $[O\,\textsc{ii}]$(b, a) doublets.}
\captionsetup{justification=centerfirst,hangindent=8pt,singlelinecheck=false,position=bottom,font={rm,md,up}}

\begin{figure}\setcounter{subfigure}{1}
\begin{center}
\label{figure.gauss}
\includegraphics[page=1, width=.45\textwidth]{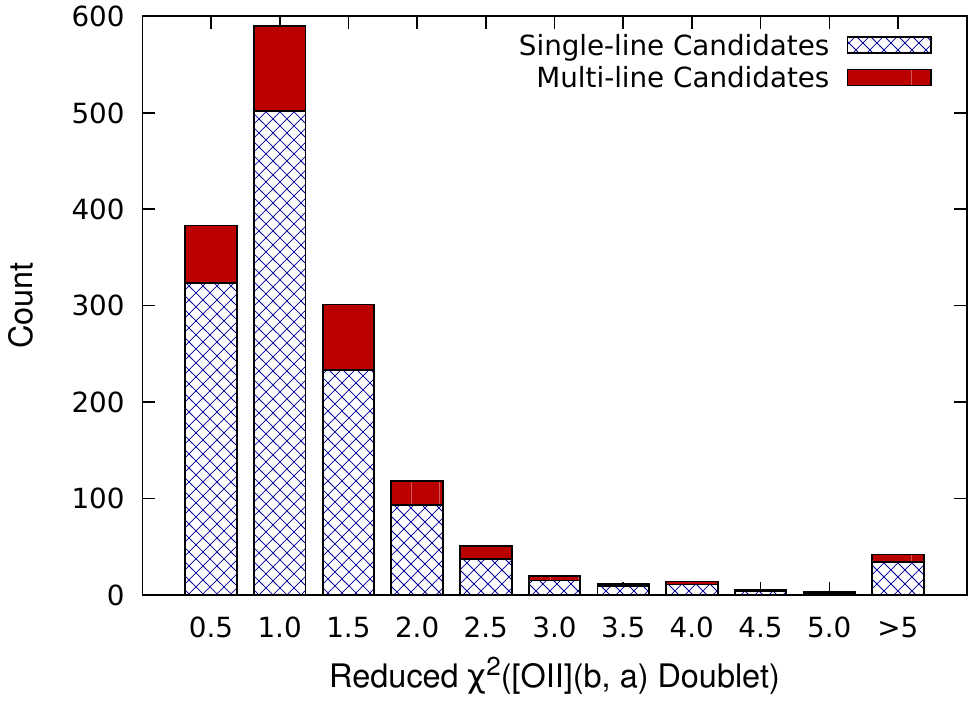}
\end{center}
\caption{\gausslinecaption}
\end{figure}

\subsection{Refining Pre-Inspection Detections}\label{subsection:method.prerefine}
We filter detections with the same initial, pre-inspection cuts applied in the BELLS method, which reject:
\begin{enumerate}
\item Single-line detections whose candidate $[O\,\textsc{ii}]$(b, a) doublet is near an unusually high occurrence of detections in the observed-frame, which is more likely explained as an under-subtracted sky emission or un-masked instrument contamination.
\item Single-line detections whose candidate $[O\,\textsc{ii}]$(b, a) doublet is near an unusually high occurrence of detections in the rest-frame, which is more likely explained as contamination from target emission-lines that are not masked in the detection method.
\item Single-line detections located at the same wavelength as a multi-line detection within the same target.
\item Single-line detections that indicate an $[O\,\textsc{ii}]$(b, a), H${\delta}$, $[O\,\textsc{iii}]$a, or H${\alpha}$ emission-line with an SN $\geq$ 3 is present when the candidate $[O\,\textsc{ii}]$(b, a) doublet is tested as a misidentified H${\delta}$, $[O\,\textsc{iii}]$a, or H${\alpha}$ emission-line.
\item Whose test Einstein radius (computed using a fiducial velocity dispersion of 250 km\,s$^{-1}$) is less than 450 Kpc.
\end{enumerate}

These cuts reduce the number of single-line and multi-line detections from 76,889 and 49,732 to 3,730 and 2,841, respectively, which can then be manually inspected as a more manageable number.

\subsection{Visual Inspection of Spectra}\label{subsection:method.inspection}

We follow the inspection methods of the BELLS survey in our manual inspection process, as stated below:
\begin{enumerate}
\item ``Visually inspect the reduced spectra of all remaining selected systems at the position of the candidate background emission-lines, to remove obvious data-reduction/data-quality artifacts.''
\item ``Inspect neighboring fiber spectra on each plate at the wavelength of the candidate background emission to rule out cross talk from bright emission-lines in neighboring spectra, and spatially localized auroral emission not subtracted by the global sky spectrum model."
\item ``Inspect the individual exposure spectra that were co-added to give the final detection spectrum, to ensure that the candidate background emission-line feature is not present in only a single frame (as would be expected for e.g., cosmic-ray hits).''
\item ``Inspect the raw spectroscopic CCD data for all exposures at the position of $[O\,\textsc{ii}]\lambda\lambda$ 3727 emission for single-line candidates, to ensure that the detection is not related to cosmetic blemishes in the CCD.''
\end{enumerate}

We included more observables in the SILO inspection process in an effort to better assess the isolation of real emission-lines of the background galaxy, improve the rejection efficiency of false detections, and gain more insight into subtle sources of contamination. The additional methods and observables we examined are listed below, in which we clarify here that references of ``sky'' in the list below refer to atmosphere emissions. We also provide a quality of assurance that the signal of the detection is from a background galaxy by assigning a grade of assurance (see grading criteria in Table \ref{table:graded_categories}) and inspection comment. The grading scheme is intended to assign an A- or higher to detections with clarity of signal approximate to the BELLS candidates. The grade assigned to each detection is dependent on the number of candidate emission-lines observed, how clearly the candidate's signal matches the emission-line patterns of a star-forming galaxy, and how isolated is the signal from regions of potential contamination. The assigned grade represents the combined assurances from the BELLS inspection methods and the additional methods listed here:

\begin{enumerate}
\item Simultaneously examine the number, quality, SN (to assess if lensing might be magnifying the signal), patterns demonstrated, and isolation of all candidate emission-lines listed in Table~\ref{table:emline} that are observed within the spectra whether the signal to noise of each candidate emission-line is weak or strong. This additional step enables the inspector to observe additional candidate emission-lines below the detection threshold to better assess the assurances of detections, which is particularly beneficial to assess single-line detections that are relatively more prone to misidentification than multi-line detections.
\item Examine each spectrum with the \texttt{ANDMASK} (obtained from the BOSS and eBOSS surveys) to flag the proximity of known sky/calibration/instrumental artefacts to each candidate emission-line. This step highlights potential artefacts that might not be eliminated entirely from the co-added spectrum.
\item Examine each spectrum with the \texttt{ORMASK} (obtained from the BOSS and eBOSS surveys) to flag the proximity of known cosmic-rays and other issues that might alter the appearance of the signal. This step is redundant with the BELLS selection method but is useful to demonstrate that an issue is apparent.
\item Examine enhanced subtle variances in the raw CCD frame at the region of the candidate $[O\,\textsc{ii}]$(b, a) doublet within each exposure. This step revealed subtle events of contamination that are not detected within the reduction pipeline and are hard to observe in the raw frame without stretching the intensity of the image by the local sigma around the local median.
\item Compare the proximity, alignment, pattern, and relative amplitude of any candidate emission-line near or over sky emissions to determine how likely each candidate emission-line may be explained as contamination from under-subtraction of sky emissions. We also compare the uncalibrated and calibrated sky spectra to reveal regions of unusual sky calibration. Trends observed in the patterns of sky contamination enable us to better calibrate the inspection process during the re-evaluation of the candidates (see Section \ref{subsection:method.iteration}).
\item If we suspected a reduction issue, we compared how the signal of the detection translates between the raw CCD frame, the individual exposures, and the co-added spectra to learn how potential contamination may be introduced during data reduction. Observed trends in the indications of reduction contamination enable us to better calibrate the inspection process during the re-evaluation of the candidates (see Section \ref{subsection:method.iteration}). 
\item Examine the calibration vector to notice any calibration variances at the location of each candidate emission-line.
\item Examine the wavelength dispersion of the spectra located near the candidate to observe any variances that can impact the measurement of the flux. 
\item Examine if the signal is near an unusually high occurrence of either detections or detections that passed the first iteration of inspection (see Section \ref{subsection:method.iteration}), which cannot be justified by relatively lower noise in the region caused by less intense or fewer sky emissions. This step helps the inspector identify regions of known and unknown contamination that are too insignificant to be filtered in the pre-inspection cuts.
\item Examine the proximity of all candidate emission-lines in Table~\ref{table:emline} to 172 emission-lines observed in the spectra of galaxies, Active Galactic Nuclei (AGN), and Quasi-Stellar Objects (QSOs)~\citep{1987ApJ...318..145F, 1985ApJ...297..166O, 2011MNRAS.414.3360R, 2008ApJ...678L..13K, 1992A&A...266..117A, 2011ApJ...740...85W, 2004A&A...421..539I, 2001A&A...378L..45I, 1998MNRAS.293L..52P, 2005ApJS..161..240T, 2004ApJ...611..107L, 2000AJ....120..562T, 1997ApJ...475..469Z, 2001AJ....122..549V, 1982ApJ...261...64O, 1999A&A...341..399S, 1996PASP..108..183O, Dhanda_2007, 1990ApJ...352..561O, 1988ApJ...330..751P}, which are recorded within the NIST Atomic Spectra Database~\citep{NIST_ASD}\footnote{NIST Atomic Spectra Database: \url{https://www.nist.gov/pml/atomic-spectra-database}}, with the list of emission-lines organized online here\footnote{Table of target emission-lines selected by Drew Chojnowski: \url{http://astronomy.nmsu.edu/drewski/tableofemissionlines.html}}. Candidate emission-lines that exist within $\approx\pm4$ Angstroms of a target emission-line are highly suspected if either the relative occurrence of detections or passing detections (see Section \ref{subsection:method.iteration}) is relatively high nearby, the width of a potentially affected $[O\,\textsc{ii}]$(b, a) imply a single emission-line, the target emission-line is known to be a source of contamination, or the region demonstrates other suspicious features identified from the first inspection.
\item Globally assess the trending quality and patterns observed within the signal revealed by the Gaussian fit information to alert the inspector to detections that contain either promising or questionable features.
\item Use the Gaussian information to observe if the shape of the $[O\,\textsc{ii}]$(b, a) doublet does not reasonably fit the $[O\,\textsc{ii}]$(b, a) Gaussian doublet. Candidate $[O\,\textsc{ii}]$(b, a) doublets that have a large reduced $\tilde{\chi}^2$ are suspected as false-positives since the signal may lack the shape of an $[O\,\textsc{ii}]$(b, a) doublet.
\item Use the Gaussian information to observe if the $[O\,\textsc{ii}]$(b, a) doublet fits an $[O\,\textsc{ii}]$(b, a) doublet Gaussian worse than a single Gaussian.
\item Use the Gaussian information to flag if the full width at half-maximum (FWHM) of the $[O\,\textsc{ii}]$(b, a) doublet is less than the separation between the emission-lines. Candidate $[O\,\textsc{ii}]$(b, a) doublets whose FWHM are less than the separation between the $[O\,\textsc{ii}]$(b, a) emission-lines are rejected as false positives if no other assurances are present.
\item Examine the candidate emission-line overlayed with the Gaussian fit model (see \ref{subsection:method.gauss}). This enabled the inspector to compare the candidate $[O\,\textsc{ii}]$(b, a) doublet with how the candidate would be approximately shaped by the redshift-dependent separation between the emission-lines if the evaluated sigma of the model is typical of emission-line candidates.
\item Highlight emission-lines other than the $[O\,\textsc{ii}]$(b, a) doublet that demonstrate patterns expected of star-forming galaxies, such as detections with an $[O\,\textsc{iii}]$a to $[O\,\textsc{iii}]$b signature ratio of 3:1 in amplitude.
\item Re-examine suspected regions and trends learned during the first iteration of inspection (see Section~\ref{subsection:method.iteration}).
\end{enumerate}

\subsection{Inspection of Lensing in Low-Resolution Images}\label{subsection:method.imaging}

Detections that pass spectroscopic inspection are next examined for interfering objects and misidentified targets within the low-resolution images. The available images used are from the Legacy survey of SDSS-I/II and the Dark Energy Camera Legacy Survey (DECaLS), the Beijing-Arizona Sky Survey~\citep[BASS;]{Zou_2017}, and the Mayall z-band Legacy Survey (MzLS) survey within the Dark Energy Spectroscopic Instrument survey~\citep[DESI;][]{2016arXiv161100036D, Dey_2019}. These images are obtained from the DESI sky server\footnote{DESI sky server: \url{http://legacysurvey.org/viewer}}. We reject detections that demonstrate clear misidentifications within the low-resolution images. A total of 1,263 single-line and 288 multi-line detections passed spectroscopic and imaging inspection to qualify as strong lensing candidates.

Though SILO candidates have not been confirmed by foreground subtraction from high-resolution images, we considered if indications of rings, multiple images, arcs, background galaxies within arc-seconds of the target, and possible overlapping colours of two galaxies may increase the assurances for a fair fraction of SILO detections. Inspiration for this additional method originates from the BELLS GALLERY survey, which published that 4 of 17 ($24\%$) of their confirmed lenses demonstrated lensing evidence within images of the target obtained from the Legacy survey of SDSS-I/II. However, the lensing evidence we expect to observe within SDSS images of SILO lens candidates cannot be directly equated to the BELLS GALLERY survey results since this survey used a sufficiently different spectroscopic selection method.

We can approximate the expected fraction of potential lensing features we can observe within SILO candidates to the fraction of lensing features observed in BELLS candidates since both surveys search the BOSS galaxies using the same detection method. We found $\frac{1}{5}$ of the BELLS candidates demonstrated indications of lensing, of which $\frac{2}{3}$ of these indications are observed in grade-A lenses, with one indication observed in a false-positive candidate. A search within low-resolution images of SILO candidates yielded a similar fraction of potential features of lensing (see Table~\ref{tables.images_counts}).

The greatest difficulty of assessing the assurances from lensing evidence within low-resolution images is that un-lensed background galaxies, satellite galaxies, clumps of star formation within targets, and flaws in the images may be the cause of the observed feature. Thus only the more well-shaped lensing evidence is considered to carry some assurance within our inspection process. However, we are able to define a level of rarity for each type and quality of lensing evidence by comparing the ratio of 'how often' the indication occurs within SILO candidates to `how often' the same indication occurs within 10,000 randomly selected galaxies from the BOSS and eBOSS surveys (see Table~\ref{tables.images_counts} and Table~\ref{tables.control_counts}). Any relative increase in the occurrence of the indication observed by this method is most likely explained by the higher occurrence of either lensed or un-lensed background galaxies within SILO candidates.

These results enabled us to develop a total grading scheme that adds the rarity of the type and quality of lensing evidence upon the assurances indicated within the spectra by assessing `how unlikely' the signal of a false spectroscopic detection matches patterns expected of emissions from a star-forming galaxy while demonstrating a type and quality of lensing evidence within low-resolution images of the target. Thus we derived a total grade based upon the grade assigned in spectroscopic inspection and biased by the rarity of the type and quality of a lensing evidence within each SILO candidate.

\subsection{Iterative Inspection Method}\label{subsection:method.iteration}
Differences in the applied instrument, reduction, target selection method, and many other factors can impact the types of contamination observed within the spectra. Some of the more subtle issues and hidden trends can be realised after inspecting a large sample of detections. We added multiple iterations of the inspection process to help the inspector assess the impact of trends of assuring signals, known contaminants, and regions demonstrating suspicious features or relations to potential issues. Several examples of contaminants learned in the first inspection are: quasars misidentified as a galaxy and at a higher redshift than catalogued, the OI emission-line misidentified as an $[O\,\textsc{ii}]$(b, a) doublet, and subtle differences between the sky emission and sky model can lead to significant residuals if the sky emission is strong. We also recorded the occurrence of passed detections per wavelength in both the rest-frame and observed-frame from the first inspection, which can help highlight regions of unexpected or underestimated contamination for the next cycle of inspection. This information enables the inspector to improve the grade of assurance assigned to the candidate during the final inspection process.

\section{Results}
\label{section:data}

\subsection{Count of Strong-Lensing Candidates}\label{subsection:results.spectracount}

We detected a total of 1,551 (1,263 single-line and 288 multi-line) candidates of various grades of assurance, with grade counts per search method, listed in Table~\ref{table:graded_results}). Of these candidates, 216 are from eBOSS targets, of which 146 are ELG targets. We assigned 838 (628 single-line and 210 multi-line) candidates a spectroscopic grade of A- or higher, and 919 with a total grade of A- or higher. If the grade-A lens probability function in SILO is similar to BELLS, then we can expect to confirm as many as $\approx400$ lenses within these candidates with a spectroscopic grade of A- or higher. Four candidate lenses contain a second background source at a higher redshift than the first background source, of which most are not highly assured, but are worth investigating as possible jackpot lenses~\citep{2008ApJ...682..964B}, as they are extremely rare objects with 2 distinct Einstein Radii. At least 74 lens candidates have been previously discovered from other surveys, including 1 from the CAmbridge Sloan Survey Of Wide ARcs in the skY survey~\citep[CASSOWARY;][]{2009MNRAS.392..104B}\footnote{CASSOWARY: \url{https://web1.ast.cam.ac.uk/ioa/research/cassowary/}}, 36 from the BELLS survey, 2 from the Strong Lensing Legacy Survey~\citep[SL2S;][]{2012ApJ...761..170G}, 1 from~\citet{Gavazzi_2014}, 26 from the Survey of Gravitationally-lensed Objects in HSC Imaging~\citep[SuGOHI;][]{Sonnenfeld_2017}\footnote{SuGOHI database: \url{http://www-utap.phys.s.u-tokyo.ac.jp/~oguri/sugohi/}}, 3 from Lenses in KiDS~\citep[LinKS;][]{Petrillo_2017, 2019MNRAS.484.3879P}\footnote{LinKS Website: \url{https://www.astro.rug.nl/lensesinkids/}}, 3 from~\citet{Jacobs_2019}, 1 from~\citet{huang2019finding}. Of these previously discovered candidates, 32 are reported as grade-A lenses from the Master Lens Database~\citep{2012hst..prop12833M}. These previously discovered candidates are not cut from the VAC (see Section~\ref{section:vac}) to enable a complete spectroscopic scan of the lensing probability function, and to be available for comparison to other surveys.


\begin{table*}
\caption{\label{table:graded_results}Count of strong lensing candidates within each grade and line search method. Column 1 lists the grade of assurance that the candidate background galaxy is strongly lensed. Columns 2 and 3 list the count of strong lensing candidates, binned by spectroscopic grade and total grade (combining the spectroscopic with imaging grade), respectively, which contained multiple emission-lines with an $SN\geq4$. Columns 4 and 5 list the count of strong lensing candidates, binned by spectroscopic grade and total grade, respectively, which contained an $[O\,\textsc{ii}]$(b, a) doublet with an $SN\geq6$}.
\centering
\begin{tabular}{ l c c c c}
\hline
\hline
 & \multicolumn{2}{c}{Multi-line Candidates} & \multicolumn{2}{c}{Single-line Candidates} \\ {Grade} & {Spectroscopic Grade} & {Total Grade} & {Spectroscopic Grade} & {Total Grade} \\
{\scriptsize (1)} & {\scriptsize (2)} & {\scriptsize (3)} & {\scriptsize (4)} & {\scriptsize (5)} \\
\hline
A+ & 105 & 125 & 105 & 205 \\
A & 58 & 49 & 277 & 257 \\
A- & 47 & 43 & 246 & 240 \\
B+ & 16 & 16 & 138 & 134 \\
B & 15 & 13 & 95 & 90 \\
B- & 15 & 18 & 169 & 158 \\
C+ & 14 & 11 & 101 & 87 \\
C & 8 & 5 & 67 & 53 \\
C- & 10 & 8 & 65 & 39 \\
Total & 288 & 288 & 1263 & 1263 \\
\hline
\end{tabular}
\end{table*}
\begin{table*}
\caption{\label{tables.images_counts}Counts and comparison to the control sample of the lensing evidence observed within low-resolution images of the lens candidates. Column 1 lists the type of lensing evidence observed. Columns 2, 4, and 6 lists the count of clearly visible, discernible, and emergent types of lensing evidence observed in the SILO candidates, respectively. Columns 3, 4, and 7 lists the clearly visible, discernible, and emergent, respectively, ratios of the number density of the type of lensing evidence observed in the lens candidates to the number density of lensing evidence observed in 10,000 randomly selected galaxies from the BOSS and eBOSS surveys. An 'NA' implies the specified evidence of lensing was not observed within the 10,000 randomly selected galaxies.}
\centering
\begin{tabular}{ l c c c c c c}
\hline
\hline
{Lensing Evidence Indicated} & \multicolumn{2}{c}{Clearly Visible} & \multicolumn{2}{c}{Discernible} & \multicolumn{2}{c}{Emergent} \\ & {Count} & {$\frac{\rho_N\ in\ Candidates}{\rho_N\ in\ Control}$} & {Count} & {$\frac{\rho_N\ in\ Candidates}{\rho_N\ in\ Control}$} & {Count} & {$\frac{\rho_N\ in\ Candidates}{\rho_N\ in\ Control}$} \\
{\scriptsize (1)} & {\scriptsize (2)} & {\scriptsize (3)} & {\scriptsize (4)} & {\scriptsize (5)} & {\scriptsize (6)} & {\scriptsize (7)} \\
\hline
Ring & 0 & NA & 5 & 10X & 7 & 3X \\
Multiple images or arcs & 3 & NA & 7 & 15X & 19 & 26X \\
Arc & 3 & 18.4X & 34 & 8.4X & 63 & 5X \\
Possible source radially nearby the lens & 8 & 3X & 75 & 2X & 46 & 1X \\
Possible overlapping colours of two galaxies & 3 & 2X & 26 & 3X & 178 & 3X \\
Total & 17 & & 147 & & 313 \\

\hline
\end{tabular}
\end{table*}

We obtained a $20\%$ increase in the number of detections over the expectation of lenses approximated in Section~\ref{section:expectation}, which may be attributed to improvements in the final DR16 version ($v5\_13\_0$) of the eBOSS spectroscopic pipeline, and considerable improvements in our search method, as detailed in Section~\ref{section:method}.

We provide examples of candidate emission-lines with high SN along with low-resolution images of the target for multi-line lens candidates in Figure~\ref{figure.flux_multiline} and single-line lens candidates in Figure~\ref{figure.flux_oneline}. These examples are sorted by grade, limited to three spectra plots per detection (for display ergonomics), with a selection in favour of targets that demonstrate lensing evidence within the low-resolution images. The grade and comment for each example are listed in Table~\ref{table:multiline} for multi-line detections and Table~\ref{table:oneline} for single-line detections, which are located in Appendix \ref{appendix:tables}. We note that these comments may describe features not displayed in the examples which impact the assigned grade.

\newcommand{\fluxmulticaption}{Examples of multi-line lens candidates, sorted by grade and selected in favour of candidates that might demonstrate a potential feature of lensing within a low-resolution image. Left: Plot of the target spectra and the position of the candidate background emission-lines with an $SN\geq4$. The grade and redshift are displayed below the plot in green. Centre: Plot of the flux, target model, and the appropriate Gaussian plus continuum model for each candidate emission-line with an $SN\geq4$. Within these plots, we represent the spectra with a black line, the foreground model with a dashed blue line, the fitted Gaussian model with a green dash-dot line, and the positions of the candidate background emission-lines with an $SN\geq4$ with a red dotted vertical line. Masked regions of the flux are shaded with grey regions, which were not used in the generation of the background emission-line detections.} Right: A low-resolution image of the candidate lens, which might display an indication that the system may be strongly lensed. The foreground redshift is presented in green on the image. Inspection comments for these candidates are listed in Table \ref{table:multiline}.
\captionsetup{justification=centerfirst,hangindent=8pt,singlelinecheck=false,position=bottom,font={rm,md,up}}

\begin{figure*}\setcounter{subfigure}{1}
\begin{center}
\label{figure.flux_multiline}
\includegraphics[page=1, width=1\textwidth]{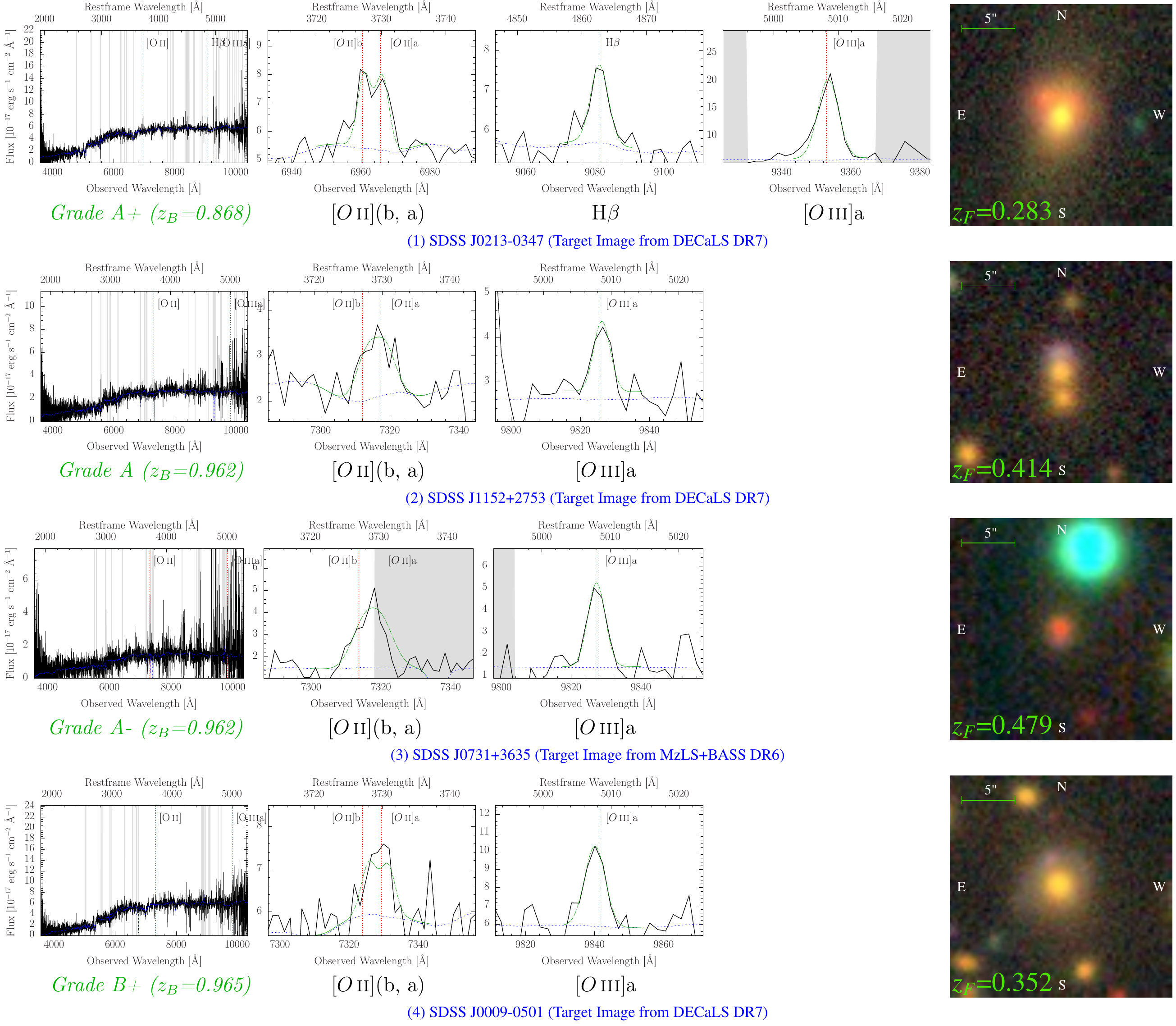}
\end{center}
\caption{\fluxmulticaption This figure is continued.}
\end{figure*}
\addtocounter{figure}{-1}

\begin{figure*}
\begin{center}
\includegraphics[page=1, width=1\textwidth]{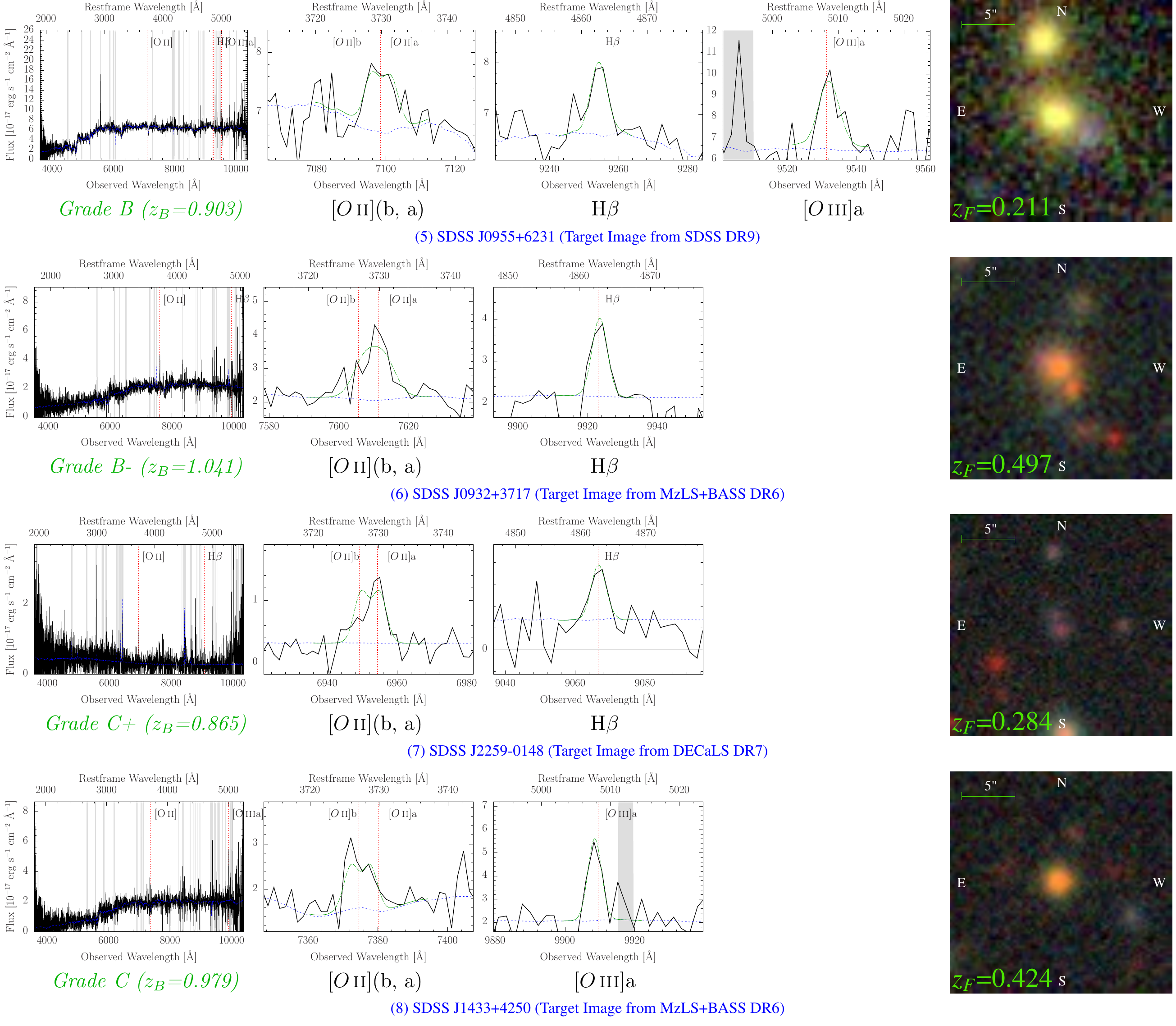}
\end{center}
\caption{\emph{Continued.} \fluxmulticaption Candidate lens SDSS J2259-0148 is an emission-line-galaxy, which confirmation of a large number of emission-line lenses can be used to constrain the mass profiles of star-forming galaxies. This figure is continued.}
\end{figure*}
\addtocounter{figure}{-1}

\setcounter{subfigure}{24}
\begin{figure*}
\begin{center}
\includegraphics[page=1, width=1\textwidth]{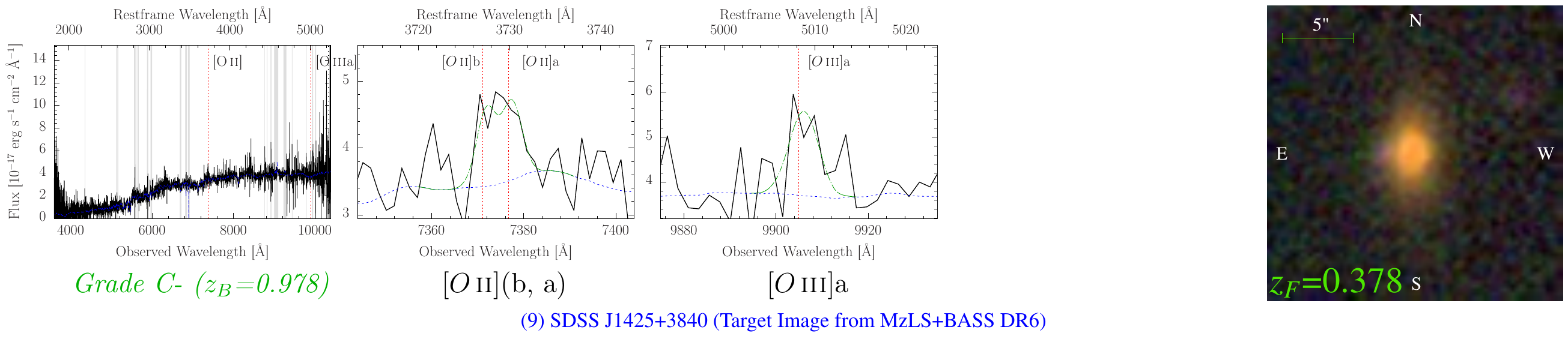}
\end{center}
\caption{\emph{Continued.} \fluxmulticaption}
\end{figure*}

\newcommand{\fluxonelinecaption}{Examples of single-line lens candidates, sorted by grade and selected in favour of candidates that might demonstrate a potential feature of lensing within a low-resolution image, but not required. Left: Plot of the target spectra and the position of the candidate background emission lines with an $SN\geq6$. The grade and redshift are displayed below the plot in green. Centre: Plot of the flux, target model, and the $[O\,\textsc{ii}]$(b, a) doublet Gaussian plus continuum model. Within these plots, we represent the spectra with a black line, the foreground model with a dashed blue line, the fitted Gaussian model with a green dash-dot line, and the positions of the candidate background emission-lines with an $SN\geq4$ with a red dotted vertical line. Masked regions of the flux are shaded with grey regions, which were not used in the generation of the background emission-line detections.} Right: A low-resolution image of the candidate lens, which might display an indication that the system may be strongly lensed. The foreground redshift is presented in green on the image. Inspection comments for these candidates are listed in Table \ref{table:oneline}.

\captionsetup{justification=centerfirst,hangindent=8pt,singlelinecheck=false,position=bottom,font={rm,md,up}}

\begin{figure*}\setcounter{subfigure}{1}
\begin{center}
\label{figure.flux_oneline}
\includegraphics[page=1, height=.86\textheight]{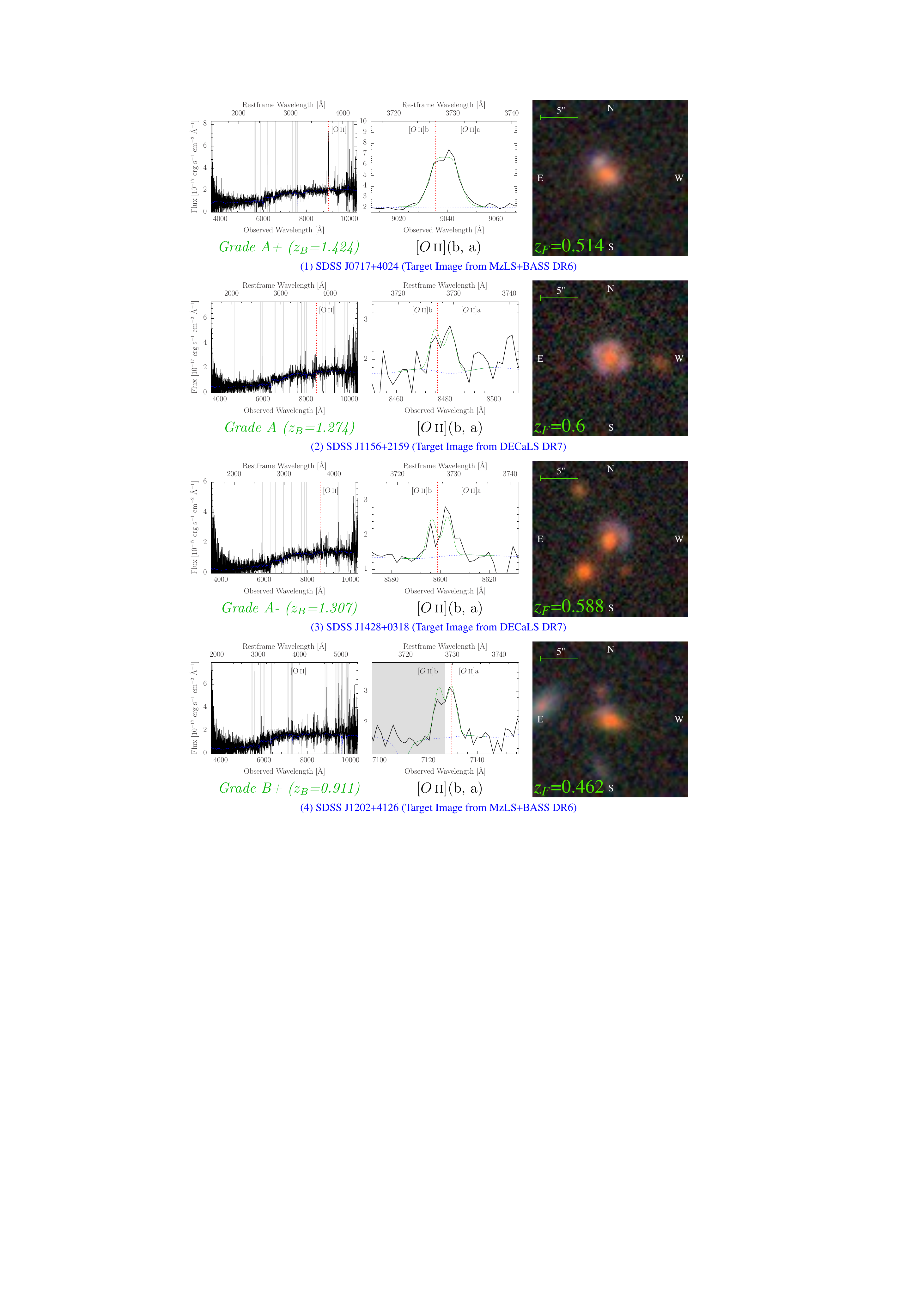}
\end{center}
\caption{\fluxonelinecaption This figure is continued.}
\end{figure*}
\addtocounter{figure}{-1}
  
\setcounter{subfigure}{12}
\begin{figure*}
\begin{center}
\includegraphics[page=1, height=.86\textheight]{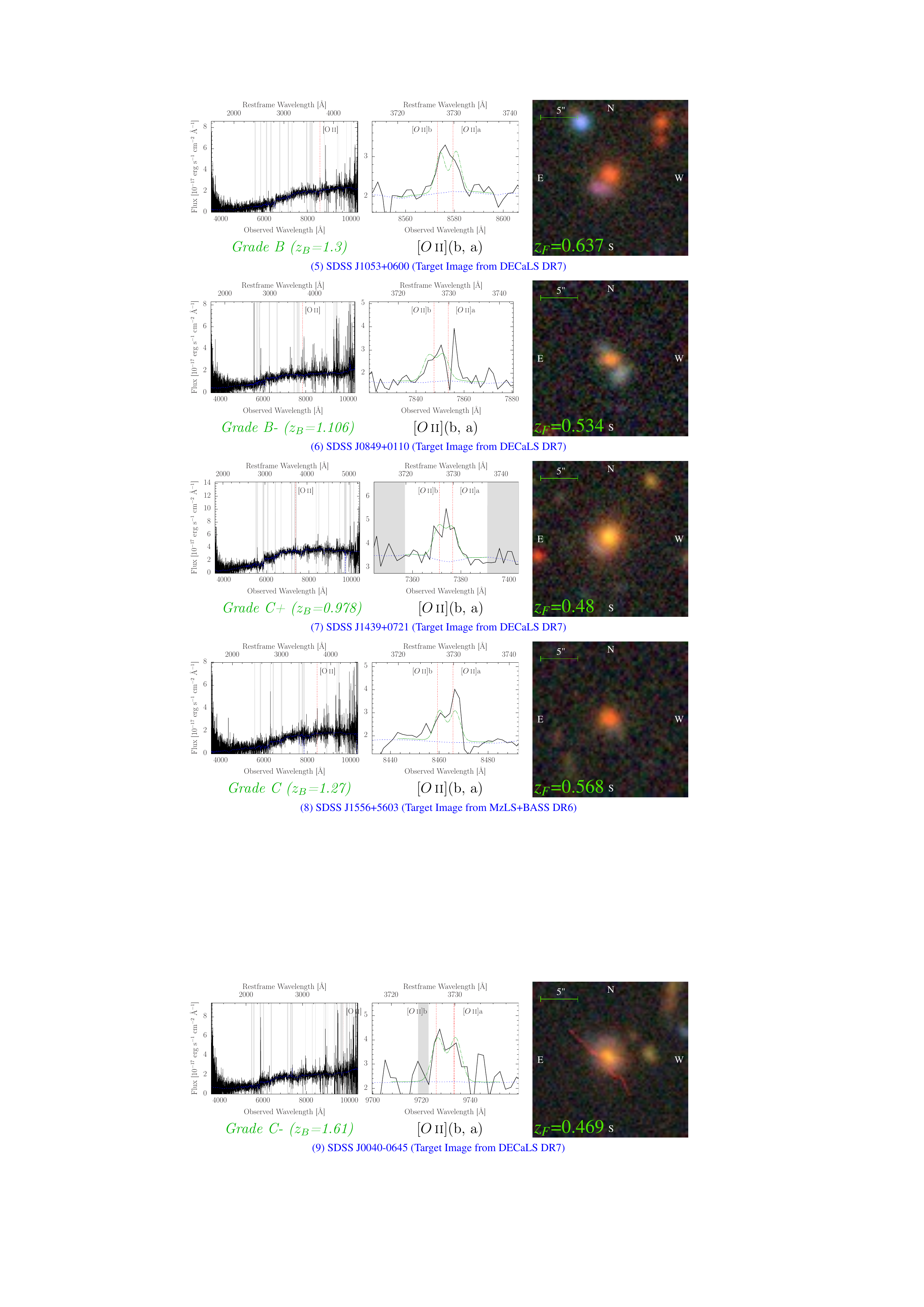}
\end{center}
\caption{\emph{Continued.} \fluxonelinecaption This figure is continued.}
\end{figure*}
\addtocounter{figure}{-1}

\setcounter{subfigure}{24}
\begin{figure*}
\begin{center}
\includegraphics[page=1, height=.215\textheight]{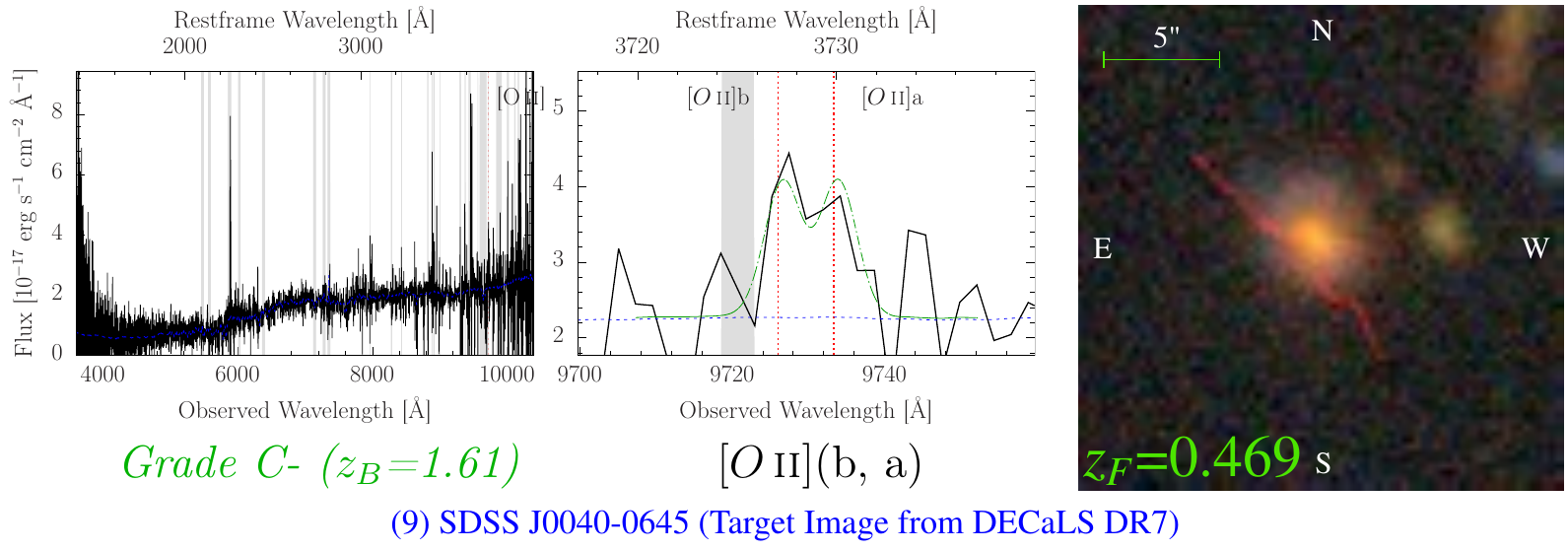}
\end{center}
\caption{\emph{Continued.} \fluxonelinecaption}
\end{figure*}

\subsection{Indications of Lensing in the Low-Resolution Images of the Candidates}\label{subsection:results.imagecount}

We observed 477 indications of lensing within low-resolution images of $\frac{1}{3}$ of the lens candidates. Examples of each type and quality of lensing evidence observed in SILO candidates are presented in Figure~\ref{figure.images}, in which these images are obtained from the DESI survey. Table~\ref{tables.images_counts} lists the counts of each type of indicated lensing feature. At least $\frac{1}{3}$ of the identified lensing features include visible warping of the source image.

\newcommand{\lensingcaption}{Examples of the type and quality of potential features of lensing within low-resolution images of strong lens candidates. Columns left to right display one clearly visible, one discernable, and one emergent indication of a specific type of potential feature of lensing, sorted from top to bottom for ring, multiple images, arc, possible source radially nearby the lens, and mixed colours of the possible overlap of the source and lens. Red circles are placed to indicate possible rings or lensed images while red arcs are placed near possible source arcs.}
\captionsetup{justification=centerfirst,hangindent=8pt,singlelinecheck=false,position=bottom,font={rm,md,up}}

\begin{figure*}\setcounter{subfigure}{1}
\label{figure.images}
\noindent
\begin{tikzpicture}
\begin{scope}[xshift=0]
\node[anchor=south east,inner sep=0] (image) at (0,0) {\includegraphics[width=1\textwidth]{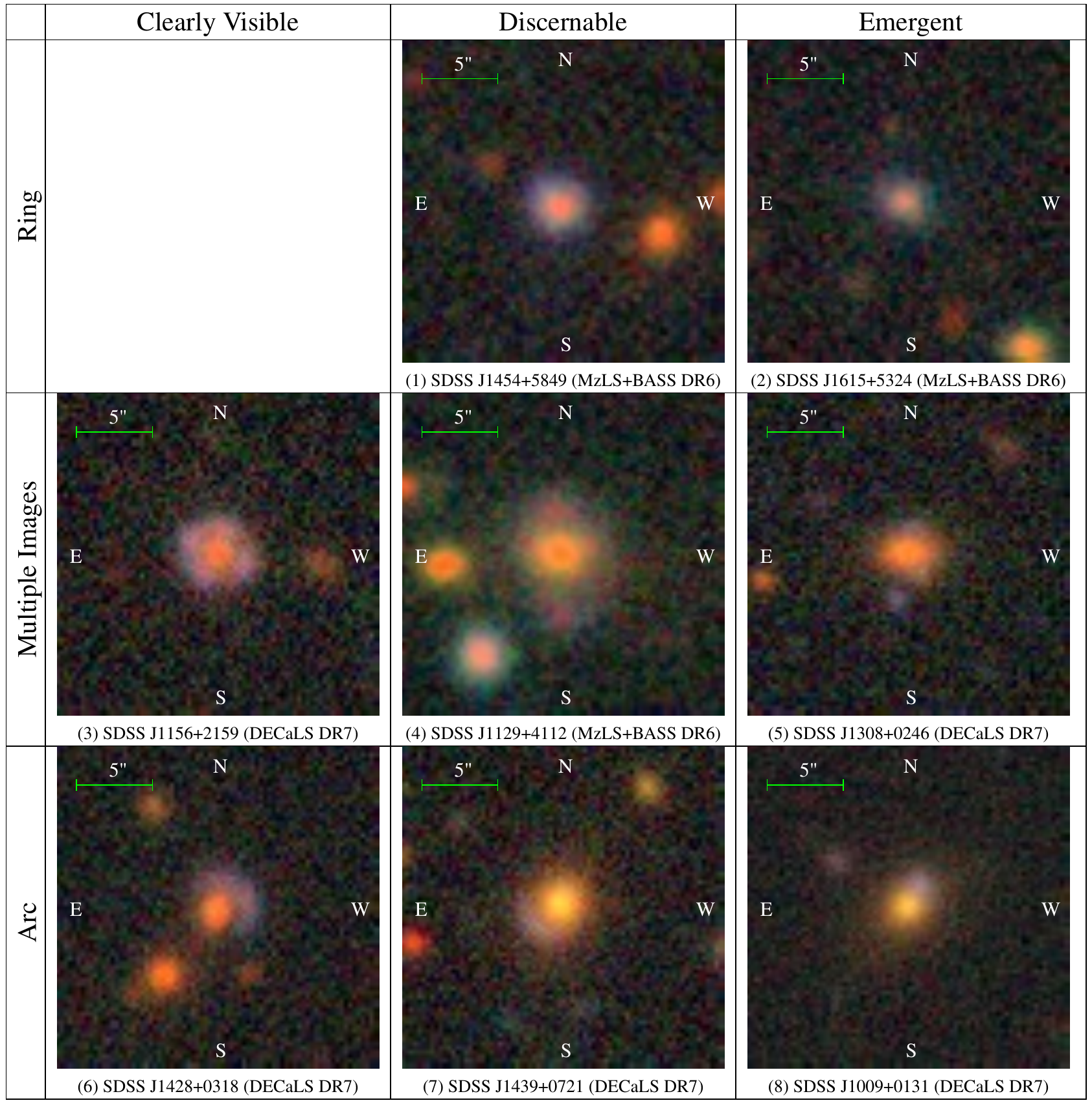}};
\begin{scope}[x={(image.south east)},y={(image.north west)}]
\draw [red] (-243pt,225pt) circle (20pt);
\draw [red] (-247pt,275pt) circle (20pt);
\draw [red] (-89pt,232pt) circle (20pt);
\draw [red] (-81pt,268pt) circle (20pt);
\draw [red] (-387.5pt,269pt) arc (40:90:20pt);
\draw [red] (-386pt,237pt) arc (-50:0:20pt);
\draw [red] (-421pt,241pt) arc (215:265:20pt);
\draw [red] (-419pt,272pt) arc (128:177:20pt);
\draw [red] (-244pt,414.5pt) circle (25pt);
\draw [red] (-85pt,417pt) circle (25pt);
\draw [red] (-381pt,77pt) arc (-30:100:25pt);
\draw [red] (-271pt,92pt) arc (180:260:25pt);
\draw [red] (-65.5pt,102.5pt) arc (30:65:25pt);
\end{scope}
\end{scope}
\end{tikzpicture}
\caption{\lensingcaption\ Strong features indicating probable strong lensing are presented in this first set of images. This figure is continued.}
\end{figure*}
\addtocounter{figure}{-1}

\setcounter{subfigure}{12}
\begin{figure*}
\noindent
\begin{tikzpicture}
\begin{scope}[xshift=0]
\node[anchor=south east,inner sep=0] (image) at (0,0) {\includegraphics[width=1\textwidth]{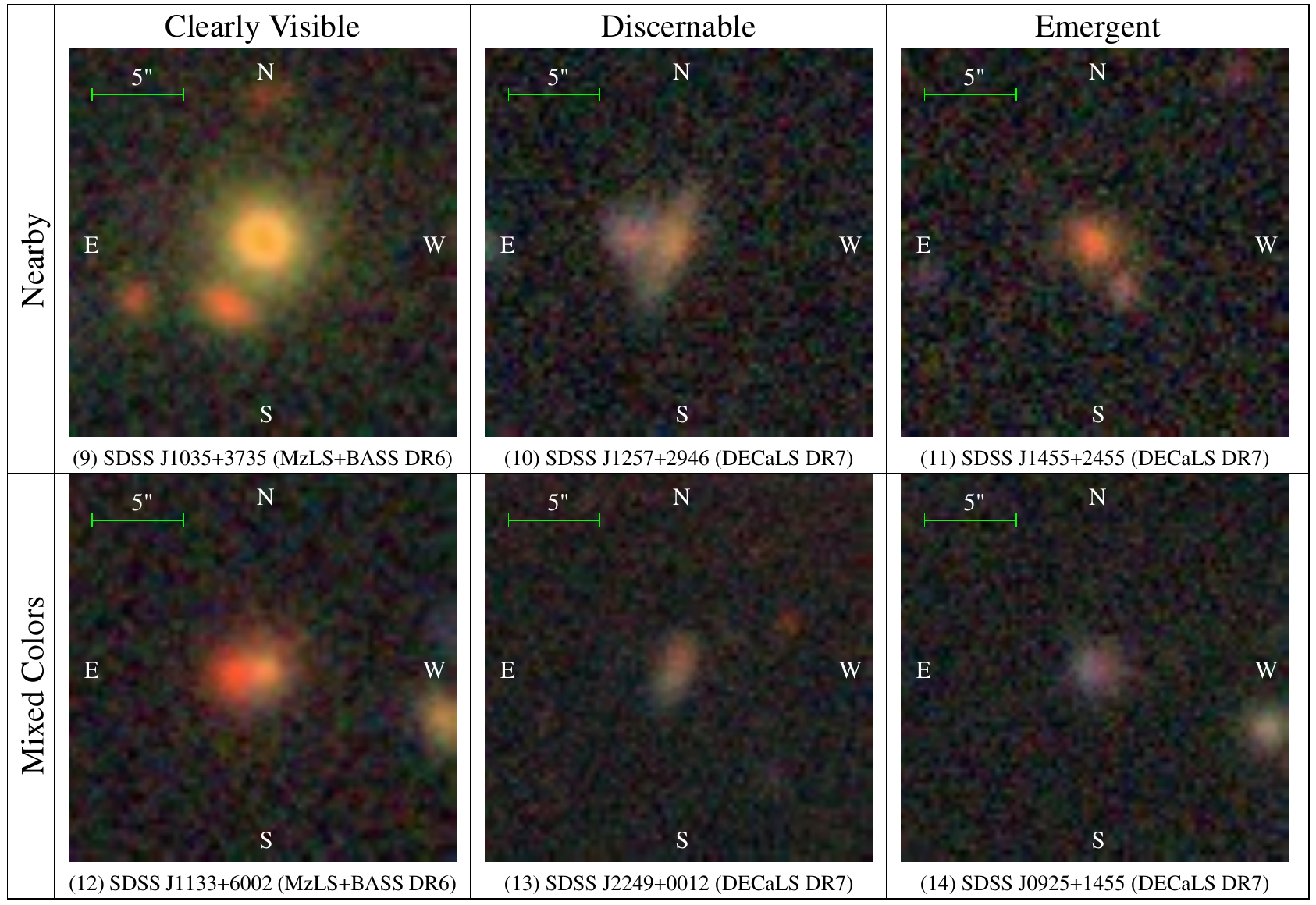}};
\begin{scope}[x={(image.south east)},y={(image.north west)}]
\draw [red] (-72.5pt,235pt) circle (10pt);
\draw [red] (-263pt,257.5pt) circle (20pt);
\draw [red] (-417pt,228.5pt) circle (15pt);
\end{scope}
\end{scope}
\end{tikzpicture}
\caption{\emph{Continued.} \lensingcaption\ Weaker features indicating possible strong lensing are presented in this figure.}
\end{figure*}

\subsection{Comparisons of SILO to the BELLS Survey}\label{subsection:results.bells}

\newcommand{\histcaption}{Comparison of the fractional distribution of the target (top) and detection redshifts (bottom) between the SILO candidates, the candidates of the BELLS survey, and the confirmed lenses of the BELLS survey.}
\captionsetup{justification=centerfirst,hangindent=8pt,singlelinecheck=false,position=bottom,font={rm,md,up}}

\begin{figure}\setcounter{subfigure}{1}
\begin{center}
\label{figure.histograms}
\includegraphics[page=1, width=.45\textwidth]{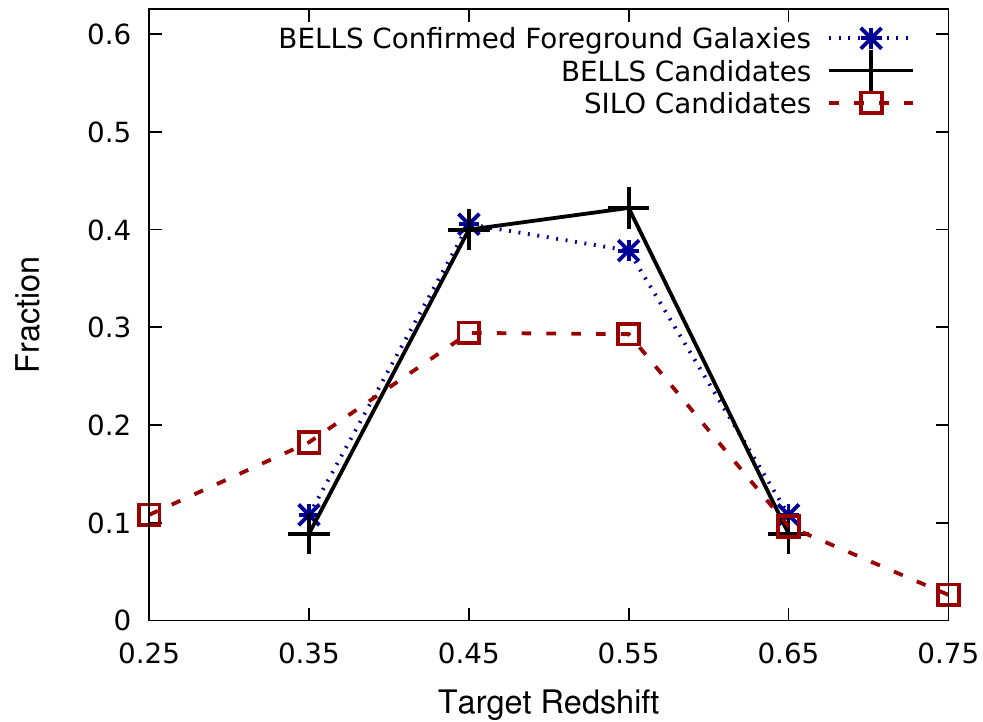}
\includegraphics[page=1, width=.45\textwidth]{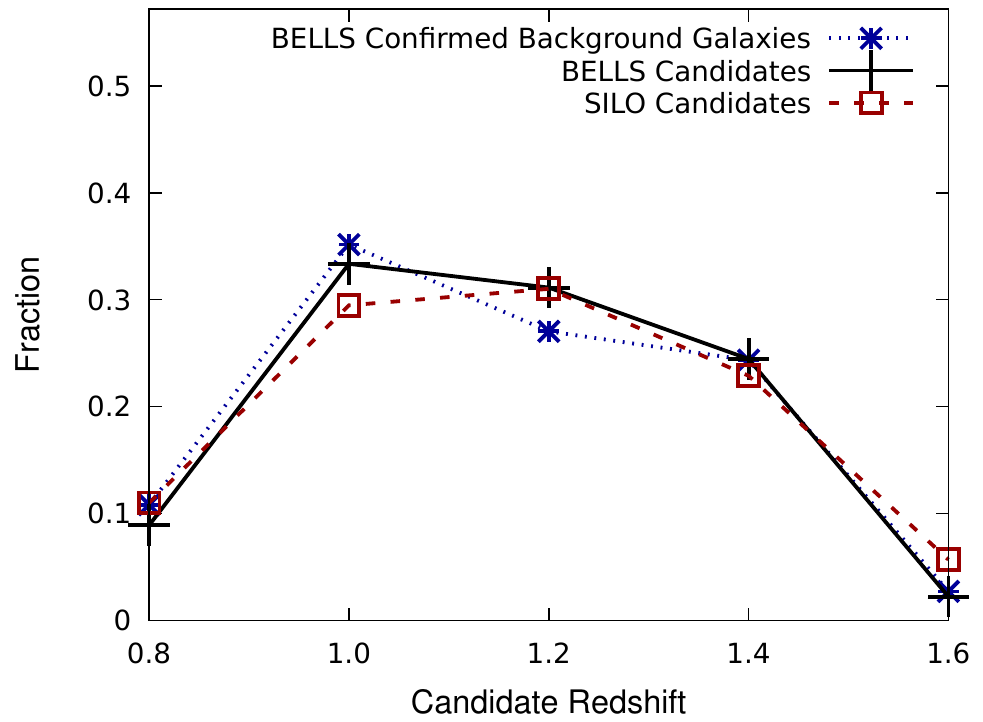}
\end{center}
\caption{\histcaption}
\end{figure}

The search for lenses in these latest reductions yielded 16.4X the lens candidates detected over the BELLS survey with quality of signal approximate to the candidates from the BELLS survey. Since the BELLS survey searched the first $\frac{1}{13}$ of the BOSS survey, the fractional increase of our candidates reveals a $\frac{1}{4}$ increase in the detection rate of SILO. We also recovered 36 of the 45 BELLS candidates. The difference is primarily caused by two effects, due to which 5 grade-A lenses were not recovered.  The first factor is that the BELLS detections were based on an earlier version of the pipeline from DR10, and pipeline improvements were made to optimize the foreground signal and redshift determination, which in some cases is less optimal for lens detection.  The second factor is that some candidates were rejected for being located within newly detected regions of sky and target emission-line contamination that SILO resolved due to its larger sample of detections (see Section ~\ref{subsection:method.prerefine}).  Because SILO was limited to BOSS and eBOSS spectra, based on the new reductions included in the SDSS-IV DR16, there would not be any SLACS lenses in the SILO VAC since SDSS-III has not been re-processed since DR8.

We inspected the targets of the BELLS candidate blindly to validate if our spectroscopic grades of A- or higher are similar to the quality of signal observed within the BELLS candidates. This test revealed that $76\%$ of the BELLS candidates within SILO are assigned a grade of A- or higher, $19\%$ are assigned a grade between B+ and B-, and $5\%$ are assigned a grade between C+ and C-. These results demonstrate that some ambiguity exists between the grade assigned by SILO to approximately represent the quality of signal observed within the BELLS candidates. This ambiguity may be caused by the differences in the reductions examined, the additional observations and inspection methods applied in SILO, or variances in the inspection criteria.

Figure~\ref{figure.histograms} demonstrates that the fractional redshift distribution of both the target and candidate background between SILO candidates and the BELLS candidates are similar. There is also an indication that the rejection rate of false-positives in the SILO project has improved over BELLS since the number density of candidates with lensing evidence within low-resolution images is $12\%$ higher for SILO when compared to BELLS.

\section{Presenting the Strong Lens Candidates in a Value-added Catalogue (VAC)}
\label{section:vac}

We included these lens candidates in a value-added catalogue (VAC) in the Sixteenth Data Release (DR16) of SDSS, to enable future studies. This VAC also enables the public to examine the lens candidates that best suite their interest, and to do follow-up observation.

The SILO VAC is located on the Science Archive Server\footnote{SILO VAC on SAS: \url{https://data.sdss.org/sas/dr16/eboss/spectro/lensing/silo/v5\_13\_0/v5\_13\_0/1.0.1}} (SAS) of SDSS and is divided into three types of files:
\begin{enumerate}
    \item The VAC contains a summary fits file (\texttt{silo\_eboss\_detections-1.0.1.fits}) that presents the detection, grade information, spectra, and comments derived from the manual inspection process. The extensions of the fits file are separated into survey information and emission-lines searched in the primary extension, as shown in Table~\ref{table:vac_table_primary}, detection information and data vectors of the spectra and model of the target, as shown in Table~\ref{table:vac_table_detection}, and candidate emission-line data with Gaussian fit information for each candidate emission-line with \(SN > 3\), as shown in Table~\ref{table:vac_table_emline}. The columns of this fits file are described and documented in the data model located on the SDSS server\footnote{SILO VAC Data model: \url{https://data.sdss.org/datamodel/files/EBOSS_SPECTRO_LENSING/silo/RUN2D/RUN1D/SILO_VER/silo_eboss_detections.html}}.
    \item The VAC contains plots of the candidate emission-lines with a detection threshold of \(SN > 4\)  located in an \texttt{images} folder, organized by plate, with paths of the form \texttt{images/PLATE/flux-PLATE-MJD-FIBERID-HITID.png}. We include comments based on our manual inspection in the caption at the bottom of these plots, with details of the quality of the signal observed, including possible emission-lines with \(SN < 4\) that are below the threshold of the automated detection code and are only noticed during manual inspection.
    \item The VAC contains low-resolution images of the target galaxy obtained from the Legacy survey of SDSS-I/II and the DESI surveys. We created 5\arcsec and 20\arcsec image cutouts for each target for each survey the image was obtained from. We generated SDSS-I/II Legacy image cutouts for all of our candidates. Since the footprints of the DESI imaging surveys partially overlap the locations of BOSS and eBOSS targets, we successfully generated DECaLS (DR7) image cutouts for 1,113 candidates and MzLS+BASS DR6 image cutouts for 532 candidates.  Note that some of the candidates have images from multiple imaging surveys.  These files are located in the \texttt{images} folder, organized by plate, with filename prefixes corresponding to \texttt{sdss\_image}, \texttt{decals\_image}, or \texttt{mzls\_bass\_image}, with either a \texttt{-5.png} or \texttt{-20.png} file extension for the 5\arcsec and 20\arcsec cutouts, respectively.
\end{enumerate}
\section{Conclusion and Discussion}\label{section:conclusion}
\subsection{Summary}\label{subsection:conclusion.summary}

We extended the Spectroscopic Identification of Lensing Objects (SILO) project to search the $\approx2$ million galaxy spectra contained within the BOSS and eBOSS data of SDSS DR16. Our spectroscopic detection algorithm is based upon the method applied in the BELLS survey and includes manual inspection grades, best-fit Gaussian distributions to background emission-lines, low-resolution images of the target, and other spectroscopic  observables to better assess each candidate and reject more false-positives. Of the 1,551 (288 multi-line and 1,263 single-line) strong lensing candidates within SILO, 838 are approximate in quality to the 45 candidates found within the BELLS survey. The selection rate of SILO candidates within the BOSS survey is $\approx\frac{1}{4}$ higher than the selection rate in the BELLS survey, which may be attributed to the search for lens candidates in the improved reductions of the BOSS data in DR16, the additional information used in the manual inspection process, and changes in the inspection criteria.

Though these strong lens candidates have not been observed by high-resolution imaging, 477 of the candidates demonstrate lensing evidence found within the low-resolution images from the Legacy survey of SDSS-I/II and the DESI Legacy survey. The fraction of potential features of lensing within SILO candidates is $12\%$ higher than observed within the BELLS candidates.

We produced a value-added catalog (VAC) for the mini Data Release 16 of SDSS-IV, as described in Section~\ref{section:vac}, which includes plots of the candidate emission-lines, low-resolution images of the target, and a summary fits file containing data for the lens candidates. The 74 lens candidates previously cited by previous surveys are referenced within the VAC.

\subsection{Discussion of How Confirmed SILO Lenses can Contribute to Galaxy Formation Studies and Future Lens Searches}\label{subsection:conclusion.examples}

Assuming a BELLS similar detection rate of grade-A lenses, we anticipate that $\approx400$ grade-A lenses can be confirmed from the lens candidates with a spectroscopic grade of A- or higher. Thus follow-up high-resolution imaging of SILO candidates may confirm over twice the number of known lenses that are spectroscopically detected within the SDSS surveys. Such a large sample of lenses can complement the results of previous surveys, enable new research objectives to be obtained, and provide resources to other search programs. We provide several examples of the applications that SILO lenses may contribute to the lensing community in the following paragraphs.

The LRG lenses found within the Legacy survey of SDSS-I/II contained target redshifts up to \(\sim 0.55\), while SILO candidates contain target redshifts up to \(1.1\) and candidate source redshifts up to \(1.713\). Thus follow-up imaging of the LRG lens candidates from SILO can extend the observations from lenses within the Legacy survey by twice the redshift, enabling farther and tighter constraints on time-dependent and time-independent properties and formation models of LRGs (see Section \ref{section:introduction}). 

Even a moderate sample of ELG lenses can significantly improve the current understanding of the mass profiles within ELGs since this area of research is less constrained. For example, the SWELLS survey spectroscopically detected 20 high-inclination disk galaxy lenses within the Legacy survey of SDSS-I/II~\cite{2011MNRAS.417.1601T} which was used to statistically verify their findings~\citep{10.1111/j.1365-2966.2012.20870.x} from a few disk lenses that a Chabrier IMF is favoured over a Salpeter IMF for disk galaxies~\citep{2011MNRAS.417.1621D, 2012MNRAS.423.1073B}. The SWELLS survey also used this sample to probe the IMF of the disk and bulge of disk galaxies~\citep{Brewer_2013}. Confirming a large sample of lenses from the 146 ELG lens candidates in SILO may provide sufficient resources to better separate and examine the mass components of the bulge, disk, and halo of spiral galaxies~\citep{1998ApJ...495..157K}.

We observe that lens searches within either images or spectra often provide resources that may be beneficial to the other search method while candidates from one search method are rarely found by the other.  The first observation that supports our reasoning is that few of the lens candidates are being detected by both imaging and spectroscopic search methods due to the physical limits of the search methods. Two recent examples of image-based search methods which affirm a fraction of our SILO detections in SDSS DR16 are:
\begin{itemize}
\item \citet{2020arXiv200504730H} report on a search for ``new'' strong gravitational lenses within the DESI Legacy Imaging Surveys, which has detected 10 SILO candidates (DESI-036.0992+02.4337, DESI-217.0936+03.3000, DESI-219.9040+07.3502, DESI-341.8329+18.0226, DESI-132.2802+22.0446, DESI-205.8655+21.7388, DESI-219.1501+21.1453, DESI-233.2341+50.1953, DESI-160.7046+62.6611, DESI-010.0606-06.7620).\\
\item The Highly Optimised Lensing Investigations of Supernovae, Microlensing Objects, and Kinematics of Ellipticals and Spirals~\citep[HOLISMOKES;][]{2020A&A...644A.163C} which has affirmed 7 SILO candidates (PS1J0737+1914, PS1J0907+4233, PS1J1233+1407, PS1J0917+3109, PS1J1134+1712, PS1J2233+3012, PS1J1105+5503).
\end{itemize}

Some image search methods subtract the lens light in images to reveal lensed features behind the relatively dominant light of the lens~\citep{Joseph_2014}. While this method can work on high-resolution images, subtraction of the lens light can remove source features within lower-resolution or lower SN images~\citep{Brault_2015}. Other methods have tried to separate or examine colour differences between the lens and source~\citep{Gavazzi_2014, Maturi_2014}. The most recently applied image searches methods typically use machine learning to search for potential source features~\citep{10.1111/j.1365-2966.2008.12880.x, 2009ApJ...694..924M, Ostrovski_2016, article, Petrillo_2017, 2019MNRAS.484.3879P, Avestruz_2019, huang2019finding, Jacobs_2019}. However, the only sources effectively detected by these methods are located beyond the relatively dominating and inerfereing light of the lens. Thus the detection of the source images by these methods typically requires the Einstein radius of the lens to be fairly larger than $1\arcsec$. In contrast, spectroscopically detected lenses from the SLACS, BELLS, BELLS GALLERY, and SWELLS surveys rarely demonstrate an Einstein radius significantly larger than $1\arcsec$ since the $1-1.5\arcsec$ radius of the fibre limits the solid angle the source can be effectively detected within. Thus the combined lenses from both image and spectroscopic detection methods can yield a large sample of lenses with a distribution of Einstein radii across most of a galactic profile while neither method can effectively recover the candidates obtained from applying the other method. Confirmation of a large sample of lenses found by SILO within the MaNGA, BOSS, and eBOSS surveys can populate the relatively under-represented distributions of Einstein radii across the LRG galactic profile, which populations cannot be obtained from either the lenses in the Legacy survey of SDSS or the lenses from the image search programs described here.

Our second observation is that both search methods utilise image and spectroscopic information, ranging from understanding how to develop a search method to confirmation and examination of the source and lens. Also, spectra and image combined searches are developing as large spectroscopic surveys (such as BOSS and eBOSS within SDSS-III/IV) and large low/higher-resolution imaging surveys~\citep[such as DESI or HSC-SSP;][]{2018PASJ...70S...8A} are observing many of the same targets. The SuGOHI survey has developed a spectra and image combined search method that subtracts the lens light from the images and spectra to detect background emission-lines within the BOSS spectra and lensing evidence within HSC-SSP images. The SuGOHI survey has searched at least $~\frac{1}{20}$ of the BOSS galaxy spectra and found at least 26 of the lens candidates contained within the SILO VAC. SILO has also included image information into its spectroscopic inspection process by examining low-resolution images of the spectroscopically selected candidates (see Section \ref{subsection:method.imaging}), which enabled us to compare if certain types and quality of lensing evidence occur more often in spectroscopically selected candidates (see Table~\ref{tables.images_counts}) than in a random sample of BOSS and eBOSS galaxies (see Table~\ref{tables.control_counts}). The image information enabled us to improve the assurances for $\frac{1}{3}$ of our candidates and may provide information to lens searches within low-resolution images about large-scale trends of lensing evidence supported by spectroscopic detection. Confirmation of SILO lens candidates can test if these trends have merit in any future image or spectroscopic searches within DESI and SDSS.

\section*{Acknowledgements}

Funding for the Sloan Digital Sky 
Survey IV has been provided by the 
Alfred P. Sloan Foundation, the U.S. 
Department of Energy Office of 
Science, and the Participating 
Institutions. 

SDSS-IV acknowledges support and 
resources from the Center for High 
Performance Computing  at the 
University of Utah. The SDSS 
website is www.sdss.org.

SDSS-IV is managed by the 
Astrophysical Research Consortium 
for the Participating Institutions 
of the SDSS Collaboration including 
the Brazilian Participation Group, 
the Carnegie Institution for Science, 
Carnegie Mellon University, Center for 
Astrophysics | Harvard \& 
Smithsonian, the Chilean Participation 
Group, the French Participation Group, 
Instituto de Astrof\'isica de 
Canarias, The Johns Hopkins 
University, Kavli Institute for the 
Physics and Mathematics of the 
Universe (IPMU) / University of 
Tokyo, the Korean Participation Group, 
Lawrence Berkeley National Laboratory, 
Leibniz Institut f\"ur Astrophysik 
Potsdam (AIP),  Max-Planck-Institut 
f\"ur Astronomie (MPIA Heidelberg), 
Max-Planck-Institut f\"ur 
Astrophysik (MPA Garching), 
Max-Planck-Institut f\"ur 
Extraterrestrische Physik (MPE), 
National Astronomical Observatories of 
China, New Mexico State University, 
New York University, University of 
Notre Dame, Observat\'ario 
Nacional / MCTI, The Ohio State 
University, Pennsylvania State 
University, Shanghai 
Astronomical Observatory, United 
Kingdom Participation Group, 
Universidad Nacional Aut\'onoma 
de M\'exico, University of Arizona, 
University of Colorado Boulder, 
University of Oxford, University of 
Portsmouth, University of Utah, 
University of Virginia, University 
of Washington, University of 
Wisconsin, Vanderbilt University, 
and Yale University.

\section*{Data Availability}

The data described in this paper are available to the public at
\url{https://data.sdss.org/sas/dr16/eboss/spectro/lensing/silo/v5_13_0/v5_13_0/1.0.1}
and are described at
\url{https://www.sdss.org/dr16/data_access/value-added-catalogs/?vac_id=eboss-strong-gravitational-lens-detection-catalog}.
\bibliographystyle{mnras}
\bibliography{main}
\clearpage
\appendix
\section{Optimization of Gaussian Fits}\label{appendix:gaussfits}

SILO optimizes Gaussian fitting to the residual flux by implementing a gridded parameter search to find the grid with the lowest $\tilde{\chi}^2$, set the limits of the next grid search within the boundaries of the chosen grid, and repeat this process until the parameters of the fit do not vary significantly between iterations. While this method quickly converges on a fit, the initial range of possible values for each parameter must encompass the parameters of the global minimum and not be so extreme that the $\tilde{\chi}^2$ map created by the optimizer is poorly resolved. Thus we must provide the optimizer with a reasonable range of spectra to set the initial search range. 

We set the range of the initial grid search based upon the information contained within the search algorithm applied by the BELLS survey and the features demonstrated by the signal. We used a test sigma ($\sigma_{\rm test}$) from the BELLS survey of 2.4 pixels as a 'ruler' used in selecting samples of spectra and evaluating the range of most parameters used in the grid search. Since the pixels of BOSS spectra are $\approx1\angstrom$ in size, $\sigma_{\rm test}$ also approximates the typical sigma observed in single emission-lines of confirmed background galaxies within the BELLS survey.

We select a sample of the wavelength vector, residual flux, and inverse variance within 4 $\sigma_{\rm test}$ of the location of a candidate emission-line determined by the detection program. We select twice this range of data for any candidate $[O\,\textsc{ii}]$(b, a) doublet since the width of an $[O\,\textsc{ii}]$(b, a) is approximately twice a single emission-line.

Our choice of limits on each parameter is calibrated based on trail fits followed by visual inspection of the quality and robustness of the Gaussian fit. We strictly set the initial search range of wavelengths within one $\sigma_{\rm test}$ of the expected location of the candidate emission-line to prevent the optimizer from fitting any noisy features nearby when the SN of the candidate emission-line is low. We set the initial search range of amplitudes within three times the difference in the residual flux, centred on the maximum value of the residual flux, and evaluated either within one $\sigma_{\rm test}$ of the location of a single emission-line or one $\sigma_{\rm test}$ plus the spacing between the $[O\,\textsc{ii}]$(b, a) emission-lines. The maximum sigma the optimizer can search is set to twice the sigma approximated from the shape of the emission-line within the residual flux. 

To set the initial range in searchable heights, we first find the local median in the residual flux located either between 5 and 15 $\sigma_{\rm test}$ of the location a single emission-line, or twice these ranges for a candidate $[O\,\textsc{ii}]$(b, a) doublet. We then set the initial range in searchable heights around twice the median of the median. Performance tests of the search range on the fit parameters revealed that we also must constrain the height to be positive as to prevent a negative bias in the fitted height caused by differences between the Gaussian model and semi-Gaussian signal.

\newpage
\section{Tables}\label{appendix:tables}
\begin{table*}
\caption{\label{table:graded_categories}Spectroscopic grading criteria used in the inspection process. Column 1 lists the grades a candidate can be assigned. Column 2 lists the criteria for each grade. Grades can be appended with a sign of $+$ or $-$ to indicate if the assurances within the spectra are more or less than the stated criteria.}
\centering
\begin{tabular}{ l l }
\hline
\hline
{Grade} & {Criteria} \\
{\scriptsize (1)} & {\scriptsize (2)} \\
\hline
Grade A & \parbox{16cm}{\vspace{.2cm} The signal is judged as likely by at least two SN$\geq4$ candidate background emission-lines demonstrate one of the following:
\begin{enumerate}
\item Clear patterns of a star-forming galaxy, which any potential contamination is unlikely to interfere with the quality of the signal significantly.
\item Patterns demonstrate minor differences from the expected pattern of a star-forming galaxy and whose high SN$\approx\geq12$ dominates any potential contamination.
\item Patterns demonstrate minor differences from the expected pattern of a star-forming galaxy with additional SN$\leq4$ emission-line(s) to assure that the pattern is unlikely to be generated from any potential contamination.
\end{enumerate}
\vspace{.2cm}} \\
Grade B & \parbox{16cm}{\vspace{.2cm} The signal is judged as probable by at least two SN$\geq4$ candidate background emission-lines that demonstrate one of the following:
\begin{enumerate}
\item Patterns demonstrate no less than some difference from the patterns expected of a star-forming galaxy.
\item Patterns demonstrate little difference from the patterns expected of a star-forming galaxy that are overlapped by suspected regions of influential contamination.
\item Signal demonstrates a milder combination of the quality and contamination issues of 1 and 2.
\item Signal would have been assigned a Grade C but is reinforced by the presence of other potential emission-lines.
\end{enumerate}
\vspace{.2cm}} \\
Grade C & \parbox{16cm}{\vspace{.2cm} The signal is judged as possible by at least two SN$\geq4$ candidate background emission-lines that demonstrate one of the following:
\begin{enumerate}
\item Patterns demonstrate a significant difference from the patterns expected of a star-forming galaxy.
\item Patterns demonstrate little difference from the patterns expected of a star-forming galaxy and are located within suspected regions of dominant contamination.
\item Signal demonstrates a milder combination of the quality and contamination issues of 1 and 2.
\item The two $SN\geq4$ emission-lines are evaluated as unlikely to indicate a background galaxy but are reinforced by the presence of other potential emission-lines.
\end{enumerate}
\vspace{.2cm}} \\
\hline
\end{tabular}
\end{table*}
\begin{table*}
\caption{\label{tables.control_counts}Counts of the types and quality of potential features of lensing observed within 10,000 low-resolution images of BOSS and eBOSS targets obtained from the DESI Legacy survey, which control sample is used to assess the rarity of each potential feature of lensing. Column 1 lists the type of potential lensing feature, while columns 2 through 4 list how discernable is the feature.}
\centering
\begin{tabular}{ l c c c }
\hline
\hline
{Indication Type} & {Clearly Visible} & {Discernible} & {Emergent} \\
{\scriptsize (1)} & {\scriptsize (2)} & {\scriptsize (3)} & {\scriptsize (4)} \\
\hline
Ring & 0 & 3 & 15 \\
Multiple images or arcs & 0 & 3 & 5 \\
Arc & 1 & 25 & 77 \\
Possible source radially nearby the lens & 10 & 51 & 341 \\
Possible overlapping colours of two galaxies & 17 & 252 & 312 \\
Total & 28 & 334 & 750 \\

\hline
\end{tabular}
\end{table*}
\begin{table*}
\caption{\label{table:multiline}This table lists the System Name in Column 1, the Spectroscopic and Total Grades in Columns 2 and 3, and the Manual Inspection Comment in Column 4, for multi-line examples in Figure~\ref{figure.flux_multiline}. This table is \emph{continued}.}
\centering
\begin{tabular}{ l c c p{11.1cm} }
\hline
\hline
{System Name} & {Spectroscopic} & {Total} & {Manual Inspection Comment} \\
{\scriptsize (1)} & {\scriptsize (2)} & {\scriptsize (3)}  & {\scriptsize (4)}\\
\hline
SDSS~J0213-0347 & A+ & A+ & Well formed $[O\,\textsc{ii}]$(b, a). Shape of $[O\,\textsc{ii}]$(b, a) best matches a double gaussian. SN greater than 12 for $[O\,\textsc{ii}]$(b, a). Well formed $[O\,\textsc{iii}]$(b, a) 1/3 pattern. Multiple emission-line evidence. Emergent H${\delta}$. Emergent H${\gamma}$. Well formed H${\beta}$. Well formed $[O\,\textsc{iii}]$b. Well formed $[O\,\textsc{iii}]$a. SN greater than 12 for $[O\,\textsc{iii}]$a. Potential sky contamination near $[O\,\textsc{ii}]$(b, a). Sky emissions located near left side of candidate emission-line is approximate or less than 20 X 10$^{-17}$ ergs/s/cm$^{2}$/Ang. Nearby sky emission(s) are either relatively insignificant or misaligned to previously stated candidate emission, and thus less likely to cause a false emission-line generated from undersubtraction. Occurance of detections observed in restframe near left side of candidate emission-line is more common than the local median around $[O\,\textsc{ii}]$(b, a). Potential sky contamination near H${\delta}$. Sky emissions located near doublet is approximate or less than 20 X 10$^{-17}$ ergs/s/cm$^{2}$/Ang. Nearby sky emission(s) are either relatively insignificant or misaligned to previously stated candidate emission, and thus less likely to cause a false emission-line generated from undersubtraction. Potential sky contamination near H${\gamma}$. Sky emissions located near left side of candidate emission-line is approximate or less than 20 X 10$^{-17}$ ergs/s/cm$^{2}$/Ang. Nearby sky emission(s) are either relatively insignificant or misaligned to previously stated candidate emission, and thus less likely to cause a false emission-line generated from undersubtraction. Visual inspection of image reveal supportive indications of an arc (discernable in DECaLS image and SDSS image), which implies a total grade of A+. \\
SDSS~J1152+2753 & A & A+ & Discernable $[O\,\textsc{ii}]$(b, a). Shape of $[O\,\textsc{ii}]$(b, a) best matches a double gaussian. Multiple emission-line evidence. Emergent $[O\,\textsc{iii}]$b. Discernable $[O\,\textsc{iii}]$a. Potential sky contamination near $[O\,\textsc{ii}]$(b, a). Sky emissions located near right side of candidate emission-line is approximately between 20 X 10$^{-17}$ ergs/s/cm$^{2}$/Ang and 30 X 10$^{-17}$ ergs/s/cm$^{2}$/Ang. Nearby sky emission(s) are either relatively insignificant or misaligned to previously stated candidate eion, and thus less likely to cause a false emission-line generated from undersubtraction. Occurance of detections observed in restframe near right side of candidate emission-line is significantly more common than the local median around $[O\,\textsc{iii}]$a. Visual inspection of image reveal supportive indications of an arc (discernable in DECaLS image and SDSS image), which implies a total grade of A+. \\
SDSS~J0731+3635 & A- & A & Signal width matches $[O\,\textsc{ii}]$(b, a). Shape of $[O\,\textsc{ii}]$(b, a) best matches a mix between double and single gaussian. Multiple emission-line evidence. Emergent H${\beta}$. Well formed $[O\,\textsc{iii}]$a. Potential sky contamination near $[O\,\textsc{ii}]$(b, a). Sky emissions located near centre of candidate emission-line is approximate or greater than 30 X 10$^{-17}$ ergs/s/cm$^{2}$/Ang. Sky emissions located near left side of candidate emission-line is approximate or less than 20 X 10$^{-17}$ ergs/s/cm$^{2}$/Ang. Visual inspection of image reveal supportive indications of possible mix ed colours of background object with foreground target (emergent in MzLS+BASS image), which implies a total grade of A. \\
SDSS~J0009-0501 & B+ & A+ & Discernable $[O\,\textsc{ii}]$(b, a). Shape of $[O\,\textsc{ii}]$(b, a) best matches a mix between double and single gaussian. Multiple emission-line evidence. Discernable $[O\,\textsc{iii}]$a. Potential sky contamination near $[O\,\textsc{ii}]$(b, a). Sky emissions located near right side of candidate emission-line is approximate or greater than 30 X 10$^{-17}$ ergs/s/cm$^{2}$/Ang. Sky emissions located near left side of candidate emission-line is approximate or greater than 30 X 10$^{-17}$ ergs/s/cm$^{2}$/Ang. Candidate emission-line $[O\,\textsc{ii}]$(b, a) might be mistaken as foreground $[Fe\,\textsc{vi}]$. Occurance of detections observed in restframe near left side of candidate emission-line is significantly more common than the local median around $[O\,\textsc{ii}]$(b, a). Occurance of detections observed in restframe near right side of candidate emission-line is more common than the local median around $[O\,\textsc{ii}]$(b, a). Visual inspection of image reveal supportive indications of an arc (emergent in DECaLS image and SDSS image), which implies a total grade of A+. \\
SDSS~J0955+6231 & B & A+ & Emergent $[O\,\textsc{ii}]$(b, a). Shape of $[O\,\textsc{ii}]$(b, a) best matches a mix between double and single gaussian. Multiple emission-line evidence. Emergent H${\beta}$. Discernable $[O\,\textsc{iii}]$a. Potential sky contamination near $[O\,\textsc{ii}]$(b, a). Sky emissions located near centre of candidate emission-line is approximate or less than 20 X 10$^{-17}$ ergs/s/cm$^{2}$/Ang. Nearby sky emission(s) are either relatively insignificant or misaligned to previously stated candidate emission, and thus less likely to cause a false emission-line generated from undersubtraction. Potential sky contamination near H${\beta}$. Candidate emission-line $[O\,\textsc{iii}]$a might be mistaken as foreground $[Ar\,\textsc{i}]$. Visual inspection of image reveal supportive indications of an arc (emergent in SDSS image), which implies a total grade of A+. \\
\hline
\end{tabular}
\end{table*}

\addtocounter{table}{-1}

\begin{table*}
\caption{Continued. This table lists the System Name in Column 1, the Spectroscopic and Total Grades in Columns 2 and 3, and the Manual Inspection Comment in Column 4, for multi-line examples in Figure~\ref{figure.flux_multiline}.}~\ref{figure.flux_multiline}.
\centering
\begin{tabular}{ l c c p{11.1cm} }
\hline
\hline
{System Name} & {Spectroscopic} & {Total} & {Manual Inspection Comment} \\
{\scriptsize (1)} & {\scriptsize (2)} & {\scriptsize (3)} & {\scriptsize (4)}\\
\hline
SDSS~J0932+3717 & B- & B+ & Discernable $[O\,\textsc{ii}]$(b, a). Shape of $[O\,\textsc{ii}]$(b, a) best matches a single gaussian. Multiple emission-line evidence. Discernable H${\beta}$. Potential sky contamination near $[O\,\textsc{ii}]$(b, a). Sky emissions located near right side of candidate emission-line is approximate or less than 20 X 10$^{-17}$ ergs/s/cm$^{2}$/Ang. Sky emissions located near left side of candidate emission-line is approximate or less than 20 X 10$^{-17}$ ergs/s/cm$^{2}$/Ang. Suspicious calibration variance or sky model variance observed. Sky contamination H${\beta}$. Visual inspection of image reveal supportive indications of possible mix ed colours of background object with foreground target (emergent in MzLS+BASS image and SDSS image), which implies a total grade of B+. \\
SDSS~J2259-0148 & C+ & B- &  Discernable $[O\,\textsc{ii}]$(b, a). Shape of $[O\,\textsc{ii}]$(b, a) best matches a mix between double and single gaussian. Multiple emission-line evidence. Emergent H${\beta}$. Emergent $[O\,\textsc{iii}]$b. Discernable $[O\,\textsc{iii}]$a. Potential sky contamination near $[O\,\textsc{ii}]$(b, a). Sky emissions located near left side of candidate emission-line is approximately between 20 X 10$^{-17}$ ergs/s/cm$^{2}$/Ang and 30 X 10$^{-17}$ ergs/s/cm$^{2}$/Ang. Nearby sky emission(s) are either relatively insignificant or misaligned to previously stated candidate emission, and thus less likely to cause a false emission-line generated from undersubtraction. Occurance of detections observed in restframe near centre of candidate emission-line is significantly more common than the local median around $[O\,\textsc{ii}]$(b, a). Potential sky contamination near $[O\,\textsc{iii}]$b. Potential sky contamination near $[O\,\textsc{iii}]$a. Visual inspection of image reveal supportive indications of possible mix ed colors of background object with foreground target (emergent in DECaLS image), which implies a total grade of B-. \\
SDSS~J1433+4250 & C & C & Emergent $[O\,\textsc{ii}]$(b, a). Shape of $[O\,\textsc{ii}]$(b, a) best matches a double gaussian. Multiple emission-line evidence. Well formed $[O\,\textsc{iii}]$a. Potential sky contamination near $[O\,\textsc{ii}]$(b, a). Sky emissions located near left side of candidate emission-line is approximately between 20 X 10$^{-17}$ ergs/s/cm$^{2}$/Ang and 30 X 10$^{-17}$ ergs/s/cm$^{2}$/Ang. Evidence observed that undersubtraction influenced signal. Potential sky contamination near $[O\,\textsc{iii}]$a. Nearby sky emission(s) are either relatively insignificant or miss aligned to previously stated candidate emission, and thus less likely to cause a false emission-line generated from undersubtraction. Occurance of detections observed in restframe near left side of candidate emission-line is significantly more common than the local median around $[O\,\textsc{iii}]$a. \\
SDSS~J1425+3840 & C- & B- & Shape of $[O\,\textsc{ii}]$(b, a) best matches a mix between double and single gaussian. Multiple emission-line evidence. Emergent $[O\,\textsc{iii}]$a. Potential sky contamination near $[O\,\textsc{ii}]$(b, a). Sky emissions located near left side of candidate emission-line is approximate or greater than 30 X 10$^{-17}$ ergs/s/cm$^{2}$/Ang. Nearby sky emission(s) are either relatively insignificant or misaligned to previously stated candidate emission, and thus less likely to cause a false emission-line generated from undersubtraction. Visual inspection of image reveal supportive indications of nearby object (emergent in MzLS+BASS image and SDSS image), which implies a total grade of B-. \\
\hline
\end{tabular}
\end{table*}
\begin{table*}
\caption{\label{table:oneline}This table lists the System Name in Column 1, the Spectroscopic and Total Grades in Columns 2 and 3, and the Manual Inspection Comment in Column 4, for} single-line examples in Figure~\ref{figure.flux_oneline}. This table is \emph{continued}.
\centering
\begin{tabular}{ l c c p{11.1cm} }
\hline
\hline
{System Name} & {Spectroscopic} & {Total} & {Manual Inspection Comment} \\
{\scriptsize (1)} & {\scriptsize (2)} & {\scriptsize (3)}  & {\scriptsize (4)}\\
\hline
SDSS~J0717+4024 & A+ & A+ & Well formed $[O\,\textsc{ii}]$(b, a). Shape of $[O\,\textsc{ii}]$(b, a) best matches a double gaussian. SN greater than 12 for $[O\,\textsc{ii}]$(b, a). Potential sky contamination near $[O\,\textsc{ii}]$(b, a). Sky emissions located near centre of candidate emission-line is approximate or less than 20 X 10$^{-17}$ ergs/s/cm$^{2}$/Ang. Sky emissions located near right side of candidate emission-line is approximate or less than 20 X 10$^{-17}$ ergs/s/cm$^{2}$/Ang. Nearby sky emission(s) are either relatively insignificant or misaligned to previously stated candidate emission, and thus less likely to cause a false emission-line generated from undersubtraction. $[O\,\textsc{ii}]$(b, a) doublet feature is indicated in individual exposures. Visual inspection of image reveal supportive indications of several arcs (discernable in MzLS+BASS image and SDSS image), which implies a total grade of A+. \\
SDSS~J1156+2159 & A & A+ &Emergent $[O\,\textsc{ii}]$(b, a). Shape of $[O\,\textsc{ii}]$(b, a) best matches a double gaussian. Potential sky contamination near $[O\,\textsc{ii}]$(b, a). Sky emissions located near left side of candidate emission-line is approximate or greater than 30 X 10$^{-17}$ ergs/s/cm$^{2}$/Ang. Nearby sky emission(s) are either relatively insignificant or misaligned to previously stated candidate emission, and thus less likely to cause a false emission-line generated from undersubtraction. Visual inspection of image reveal supportive indications of a quad (well formed in DECaLS image and SDSS image), which implies a total grade of A+. \\
SDSS~J1428+0318 & A- & A+ & Discernable $[O\,\textsc{ii}]$(b, a). Shape of $[O\,\textsc{ii}]$(b, a) best matches a double gaussian. Potential sky contamination near $[O\,\textsc{ii}]$(b, a). Sky emissions located near left side of candidate emission-line is approximate or less than 20 X 10$^{-17}$ ergs/s/cm$^{2}$/Ang. Nearby sky emission(s) are either relatively insignificant or misaligned to previously stated candidate emission, and thus less likely to cause a false emission-line generated from undersubtraction. Visual inspection of image reveal supportive indications of an arc (well formed in DECaLS image and SDSS image), which implies a total grade of A+. \\
SDSS~J1202+4126 & B+ & A- & Discernable $[O\,\textsc{ii}]$(b, a). Shape of $[O\,\textsc{ii}]$(b, a) best matches a double gaussian. Occurance of pre-inspection passed detections observed in restframe near centre of candidate emission-line is slightly more common than the local median around $[O\,\textsc{ii}]$(b, a). Visual inspection of image reveal supportive indications of possible mix ed colours of background object with foreground target (emergent in MzLS+BASS image), which implies a total grade of A-. \\
SDSS~J1053+0600 & B & A & Discernable $[O\,\textsc{ii}]$(b, a). Shape of $[O\,\textsc{ii}]$(b, a) best matches a mix between double and single gaussian. Candidate emission-line $[O\,\textsc{ii}]$(b, a) might be mistaken as foreground $[Fe\,\textsc{iv}]$. Doublet signal is wider than a typical single emission-line. Visual inspection of image reveal supportive indications of an arc (discernable in DECaLS image and SDSS image), which implies a total grade of A. \\
SDSS~J0849+0110 & B- & A+ & Discernable $[O\,\textsc{ii}]$(b, a). Potential sky contamination near $[O\,\textsc{ii}]$(b, a). Sky emissions located near right side of candidate emission-line is approximate or greater than 30 X 10$^{-17}$ ergs/s/cm$^{2}$/Ang. Sky emissions located near left side of candidate emission-line is approximately between 20 X 10$^{-17}$ ergs/s/cm$^{2}$/Ang and 30 X 10$^{-17}$ ergs/s/cm$^{2}$/Ang. Visual inspection of image reveal supportive indications of several arcs (well formed in DECaLS image and SDSS image), which implies a total grade of A+. \\
SDSS~J1439+0721 & C+ & A+ & Discernable $[O\,\textsc{ii}]$(b, a). Shape of $[O\,\textsc{ii}]$(b, a) best matches a triple gaussian. Potential sky contamination near $[O\,\textsc{ii}]$(b, a). Sky emissions located near centre of candidate emission-line is approximate or greater than 30 X 10$^{-17}$ ergs/s/cm$^{2}$/Ang. Sky emissions located near left side of candidate emission-line is approximate or less than 20 X 10$^{-17}$ ergs/s/cm$^{2}$/Ang. Visual inspection of image reveal supportive indications of an arc (discernable in DECaLS image and SDSS image), which implies a total grade of A+. \\
SDSS~J1556+5603 & C & B- & Discernable $[O\,\textsc{ii}]$(b, a). Shape of $[O\,\textsc{ii}]$(b, a) best matches a mix between double and single gaussian. Potential sky contamination near $[O\,\textsc{ii}]$(b, a). Sky emissions located near right side of candidate emission-line is approximately between 20 X 10$^{-17}$ ergs/s/cm$^{2}$/Ang and 30 X 10$^{-17}$ ergs/s/cm$^{2}$/Ang. Sky emissions located near left side of candidate emission-line is approximate or less than 20 X 10$^{-17}$ ergs/s/cm$^{2}$/Ang. Nearby sky emission(s) are either relatively insignificant or misaligned to previously stated candidate emission, and thus less likely to cause a false emission-line generated from undersubtraction. Visual inspection of image reveal supportive indications of possible mix ed colours of background object with foreground target (emergent in MzLS+BASS image), which implies a total grade of B-. \\

\hline
\end{tabular}
\end{table*}

\addtocounter{table}{-1}

\begin{table*}
\caption{Continued. This table lists the System Name in Column 1, the Spectroscopic and Total Grades in Columns 2 and 3, and the Manual Inspection Comment in Column 4, for single-line examples in Figure}~\ref{figure.flux_oneline}.
\centering
\begin{tabular}{ l c c p{11.1cm} }
\hline
\hline
{System Name} & {Spectroscopic} & {Total} & {Manual Inspection Comment} \\
{\scriptsize (1)} & {\scriptsize (2)} & {\scriptsize (3)}  & {\scriptsize (4)}\\
\hline
SDSS~J0040-0645 & C- & B+ & Discernable $[O\,\textsc{ii}]$(b, a). Shape of $[O\,\textsc{ii}]$(b, a) best matches a double gaussian. Potential sky contamination near $[O\,\textsc{ii}]$(b, a). Sky emissions located near left side of candidate emission-line is approximate or greater than 30 X 10$^{-17}$ ergs/s/cm$^{2}$/Ang. Sky emissions located near right side of candidate emission-line is approximately between 20 X 10$^{-17}$ ergs/s/cm$^{2}$/Ang and 30 X 10$^{-17}$ ergs/s/cm$^{2}$/Ang. Suspicious calibration variance or sky model variance observed. Occurance of pre-inspection passed detections observed in restframe near centre of candidate emission-line is significantly more common than the local median around $[O\,\textsc{ii}]$(b, a). Visual inspection of image reveal supportive indications of a ring (emergent in DECaLS image and SDSS image), which implies a total grade of B+. \\
\hline
\end{tabular}
\end{table*}
\begin{table*}
\caption{\label{table:vac_table_primary}This table lists the headers of the primary extension of the fits file within the SILO value-added catalogue. Column 1 lists the identifier of the data. Column 2 lists the value of the header. Column 3 list the unit type. Column 4 list the format type. Column 5 lists the description of each header or table column.}
\centering
\begin{tabular}{ l l l l l }
\hline
\hline
{Name} & {Value} & {Unit} & {Type} & {Description} \\
{\scriptsize (1)} & {\scriptsize (2)} & {\scriptsize (3)} & {\scriptsize (4)} & {\scriptsize (5)} \\
\hline
PROJECT &  SILO &  & String &  Lens search project  \\
AUTHORS &  Michael Talbot &  & String &  SILO project creators  \\
INSP &  Michael Talbot &  & String &  Detection inspector  \\
SCANNED & BOSS+eBOSS & & String & SDSS surveys scanned by SILO \\
SILO\_VER &  1.0.1 &  & String &  Version of SILO project used  \\
RUN1D &  V5\_13\_0 &  & String &  BOSS/eBOSS 1D reduction version of spectra  \\
RUN2D &  V5\_13\_0 &  & String &  BOSS/eBOSS 2D reduction version of spectra  \\
RELEASE &  DR16 &  & String &  SDSS data release version  \\
SAMPLING &  LOG &  & String &  Wavelength sampling of spectra  \\
SPECTYPE &  COADDED &  & String &  Type of spectra searched  \\
OIIB &  3727.092 &  \AA & Float &  Restframe wavelength of OIIb  \\
OIIA &  3729.875 &  \AA & Float &  Restframe wavelength of OIIa  \\
HID &  4102.892 &  \AA & Float &  Restframe wavelength of HId  \\
HIC &  4341.684 &  \AA & Float &  Restframe wavelength of HIc  \\
HIB &  4862.683 &  \AA & Float &  Restframe wavelength of HIb  \\
OIIIB &  4960.295 &  \AA & Float &  Restframe wavelength of OIIIb  \\
OIIIA &  5008.239 &  \AA & Float &  Restframe wavelength of OIIIa  \\
NIIB &  6549.86 &  \AA & Float &  Restframe wavelength of NIIb  \\
HIA &  6564.614 &  \AA & Float &  Restframe wavelength of HIa  \\
NIIA &  6585.27 &  \AA & Float &  Restframe wavelength of NIIa  \\
SIIB &  6718.29 &  \AA & Float &  Restframe wavelength of SIIb  \\
SIIA &  6732.68 &  \AA & Float &  Restframe wavelength of SIIa  \\
\hline
\end{tabular}
\end{table*}

\begin{table*}
\caption{\label{table:vac_table_detection} This table list the columns within the DETECTION extension of the fits file within the SILO value-added catalogue. Column 1 lists the identifier of the data. Column 2 list the unit type. Column 3 list the format type. Column 4 lists the description of each header or table column.}
\begin{tabular}{ l l l p{7.1cm} }
\hline
\hline
{Name} & {Unit} & {Type} & {Description} \\
{\scriptsize (1)} & {\scriptsize (2)} & {\scriptsize (3)} & {\scriptsize (4)} \\
\hline
CATALOG\_ID & & Integer & Detection identifier for this catalogue  \\
EMLINE\_SCAN\_TYPE & & String & Background emission-line search mode: Single-line=OII(b, a) with SN>=6, Multi-line=2+ emission-lines with SN>=4  \\
DETECTION\_Z & & Float & Redshift of background candidate  \\
N\_EMLINES\_SN\_GE4 & & Integer & Number of emission-lines detected with SN>=4  \\
QUADATURE\_SUM\_SN\_GE3 & & Float & Quadrature sum of emission-lines with SN>=3  \\
SPECTRA\_GRADE & & String & Grade of spectra assurances of source candidate  \\
TOTAL\_GRADE & & String & Grade of spectra and image assurances of candidate  \\
COMMENT & & String & Comment of assuring/non-assuring features  \\
FIRST\_DETECTION\_FROM & & String & Identity of earlier discoverer  \\
FIRST\_DETECTION\_LG & & String & Strong lens grade from Masterlens database  \\
SDSS\_TARGET\_NAME & & String & SDSS RA+DEC name of target  \\
PLATE & & Integer & Plate number of BOSS/eBOSS observation of target  \\
MJD & & Integer & Modified Julian Day on observation of target  \\
FIBER\_INDEX & & Integer & Fibre on plate used for observation  \\
SDSS\_SURVEY & & String & SDSS survey the spectra was obtained from  \\
BESTOBJID & & String & SDSS unique database ID of (recommended) position-based photometric match based on RUN, RERUN, CAMCOl, FIELD, ID (same as SkyServer version) \\
SPECOBJID & & String & Unique database ID based on PLATE, MJD, FIBERID, RUN2D (same as SkyServer version) \\
CLASS\_NOQSO & & String & SDSS non-quasar classification of target \\
OBJCLASS & & String & SDSS classification of target \\
TARGET\_GALAXY\_TYPE & & String & SDSS classification of type of target galaxy (i.e. ELG, LRG). Variable is named SOURCETYPE in eBOSS. \\
TARGETOBJID & & String & SDSS unique database ID of targeting object based on RUN, RERUN, CAMCOl, FIELD, ID (same as SkyServer version)  \\
Z\_NOQSO & & Float & SDSS best redshift when ignoring QSO fits, recommended for BOSS CMASS and LOWZ targets; calculated only for survey='boss' spectra, not for any SDSS spectrograph data \\
ZERR\_NOQSO & & Float & Error in Z\_NOQSO redshift  \\
ZWARNING\_NOQSO & & Integer & SDSS bitmask of spectroscopic warning values; 0 means everything is OK  \\
RA & Degree &  Float & Target right ascension  \\
DEC & Degree &  Float & Target declination  \\
RA\_HMS & hour:':'' & String & Target right ascension in hour:min:sec  \\
DEC\_DMS & Degree:':'' & String & Target declination in degree:min:sec  \\
WAVE & \AA & Float (Array) & Wavelength vector of spectra from BOSS/eBOSS \\
FLUX & \fluxunits & Float (Array) & Coadded object flux from BOSS/eBOSS \\
IVAR & & Float (Array) & Inverse varience of spectra from BOSS/eBOSS \\
AND\_MASK & & Integer (Array) & and\_mask of coadded spectra from BOSS/eBOSS \\
OR\_MASK & & Integer (Array) & or\_mask of coadded spectra from BOSS/eBOSS \\
SKY & \fluxunits & Float (Array) & Coadded sky emissions from BOSS/eBOSS \\
WDISP & \fluxunits & Float (Array) & Wavelength dispersion of spectra from BOSS/eBOSS \\
MODEL\_SILO & \fluxunits & Float (Array) & Foreground model generated by SILO  \\
IVAR\_RESCALED & & Float (Array) & Rescaled inverse varience  \\
RESIDUAL\_FLUX & \fluxunits & Float (Array) & Residual flux of FLUX-MODEL\_SILO  \\
SN\_SPECTRA\_SG & & Float (Array) & Gaussian convolved signal-to-noise of spectra \\
\hline
\end{tabular}
\end{table*}

\begin{table*}
\caption{\label{table:vac_table_emline} This table list the columns within the EMISSION\_LINE\_ANALYSIS extension of the fits file within the SILO value-added catalogue. Column 1 lists the identifier of the data. Column 2 list the unit type. Column 3 list the format type. Column 4 lists the description of each header or table column.}
\begin{tabular}{ l l l p{7.1cm} }
\hline
\hline
{Name} & {Unit} & {Type} & {Description} \\
{\scriptsize (1)} & {\scriptsize (2)} & {\scriptsize (3)} & {\scriptsize (4)} \\
\hline
DETECTION\_CATALOG\_ID & & Integer & Detection identifier for this catalogue  \\
NAME & & String & Name of emission-line  \\
INDEX\_IN\_SPECTRA & & Integer & Index where emission-line is located in spectra  \\
EM\_WAVE & \AA & Flux & Wavelength of emission-line computed from rest-frame wavelength*(1+z) \\
SN & & Float & Gaussian convolved signal-to-noise of emission-line  \\
GAUSS\_FIT\_REPORTED & & Boolean & Gauss fit reported if SN>=3  \\
GAUSS\_WAVE & \AA & Float (Array) &  Wavelength(s) centre of Gaussian model with -/+ from 2.5, 97.5 quartiles from Monte-Carlo simulation  \\
GAUSS\_BASE\_HEIGHT & \fluxunits & Float (Array) &  Gaussian model base height with -/+ from 2.5, 97.5 quartiles from Monte-Carlo simulation  \\
GAUSS\_AMPLITUDE & \fluxunits & Float (Array) &  Gaussian model amplitude with -/+ from 2.5, 97.5 quartiles from Monte-Carlo simulation  \\
GAUSS\_SIGMA & \AA & Float (Array) & Gaussian model sigma with -/+ from 2.5, 97.5 quartiles from Monte-Carlo simulation  \\
RCHI2\_SAMPLE & & Float & Reduced $\tilde{\chi}^2$ of Gaussian fit to spectra sample \\
NDOF\_SAMPLE & & Integer & Degree of freedom of RCHI2\_SAMPLE \\
RCHI2\_3SIG & & Float & Reduced $\tilde{\chi}^2$ of Gaussian fit within 3 sigma of the model \\
NDOF\_3SIG & & Integer & Degree of freedom of RCHI2\_3SIG \\
SAMPLE\_SIZE & & Integer & Size of spectra sample used in Gaussian Fit \\
MODEL\_WAVE\_BASE & \AA & Float (Array) &  Wavelength base of Gaussian model  \\
GAUSS\_MODEL & \fluxunits & Float (Array) &  Gaussian model of residual flux  \\
FITTED\_RESIDUAL\_FLUX & \fluxunits & Float (Array) &  Residual flux segment used in Gaussian fit  \\
FITTED\_IVAR\_RESCALED & & Float (Array) & Rescaled inverse varience used in Gauss fit  \\
AND\_MASK & & Integer (Array) & SDSS AND\_MASK of sample of spectra  \\
OR\_MASK & & Integer (Array) & SDSS OR\_MASK of sample of spectra  \\
\hline
\end{tabular}
\end{table*}
\end{document}